\documentclass{LMCS}
\usepackage{amsfonts,latexsym, amssymb, amsmath}
\usepackage{proof,bussproofs}
\usepackage{url}
\usepackage{eepic}
\usepackage[dvips]{epsfig}
\usepackage{enumerate,hyperref}

\newcommand{\color}[2][]{}  
{}
\def\bc{\mathcal{B}}        
\def\mc{\mathcal{M}}        

\newcommand{\QQ}{\mathbb{Q}}

\def\lis#1#2{\mathsf{doors}(#1,#2)}
\def\prm#1{\mathbf{#1}}

\newcommand{\DLAL}{$DLAL$}
\newcommand{\LAL}{$LAL$}
\newcommand{\EAL}{$EAL$}
\newcommand{\LLL}{$LLL$}
\newcommand{\DLALS}{ $DLAL{\star}$ }
\def\bparam#1{par^{bool}(#1)}
\def\inparam#1{par^{int}(#1)}
\def\lift#1{\overline{#1}}
\def\coo#1{C_1[#1]}                      
\def\cot#1{C_2[#1]}                      

\def\llto{\mathbin{-\mkern-3mu\circ}}              
\def\lltensor{\otimes}

\def\unif#1#2{\mathcal{U}(#1, #2)}        

\def\bracket{\mathsf{bracket}}
\def\wbracket{\mathsf{wbracket}}
\def\Bra{\mathsf{Bracket}}
\def\Bang{\mathsf{Bang}}
\def\Scope{\mathsf{Scope}}

\def\Ltype{\mathsf{Ltype}}
\def\Adm{\mathsf{Adm}}
\def\Const{\mathsf{Const}}

\def\fCenter{{~ \vdash ~}}
\newcommand {\fli} {\Rightarrow}
\newcommand{\lf}{\ma{L}_{F}}              
\newcommand{\ldlal}{\ma{L}_{DLAL}}        
\newcommand{\ldlals}{\ma{L}_{DLAL\star}}  

\def\ci#1{\underline{#1}}                 
\def\succ{succ}                           
\def\FV#1{FV({#1})}           

\newcommand{\ma}{\mathcal}

\def\eras#1{{#1}^-}                   

\newcommand{\NN} {\mathbb{N}}
\newcommand{\ZZ} {\mathbb{Z}}

\newcommand {\fm} {\multimap}
 
\newcommand{\bs} {\mathord{!}}

\newcommand{\pp} {\mathord{?}}  \newcommand{\pa} {\mathord{\S}}
\newcommand{\pad} {\mathord{\bar{\S}}}

\newcommand{\al} {\alpha}
\newcommand{\la} {\lambda}
\newcommand{\La} {\Lambda}

\newcounter{number}

\def\doi{3 (4:10) 2007}
\lmcsheading%
{\doi}
{1--32}
{}
{}
{Jan.~\phantom{0}6, 2007}
{Nov.~15, 2007}
{}

\begin{document}

\title[Verification of Ptime reducibility for system F terms: type inference in
 DLAL]{ Verification of Ptime reducibility for system F terms:\\
 type inference in
 Dual Light Affine Logic.}

\author[V.~Atassi]{Vincent Atassi\rsuper a}	
\address{{\lsuper a}LIPN - UMR 7030, CNRS - Universit\'e Paris 13, F-93430 Villetaneuse, France}
\email{\{vincent.atassi,patrick.baillot\}@lipn.univ-paris13.fr}  
\thanks{{\lsuper a}Partially supported by projects NO-CoST (ANR, JC05\_43380), CRISS (ACI), GEOCAL (ACI)}

\author[P.~Baillot]{Patrick Baillot\rsuper a}	
\address{\vskip-6 pt}

\author[K.~Terui]{Kazushige Terui\rsuper b}	
\address{{\lsuper b}National Institute of Informatics,
Tokyo, Japan}	
\email{terui@nii.ac.jp}  
\thanks{{\lsuper b}Partially supported by Grant-in-Aid for Scientific
Research, MEXT, Japan.}	



\keywords{Linear logic, lambda calculus, implicit computational
complexity, type inference, polynomial time complexity, polymorphic
types, Light linear logic} 
\subjclass{F.4.1; F.2.2; D.1.1}

\titlecomment{}

\begin{abstract}

  In a previous work Baillot and Terui introduced Dual light affine
  logic (\DLAL) as a variant of Light linear logic suitable for
  guaranteeing complexity properties on lambda calculus terms: all
  typable terms can be evaluated in polynomial time by beta reduction
  and all Ptime functions can be represented. In the present work we
  address the problem of typing lambda-terms in second-order \DLAL.
  For that we give a procedure which, starting with a term typed in
  system F, determines whether it is typable in \DLAL\ and outputs a
  concrete typing if there exists any.  We show that our procedure can
  be run in time polynomial in the size of the original Church typed
  system F term.

 \end{abstract}

\maketitle

\section{Introduction}

Several works have studied programming languages with intrinsic
computational complexity properties. This line of research, Implicit
computational complexity (ICC), is motivated both by the perspective
of automated complexity analysis, and by foundational goals, in
particular to give natural characterisations of complexity classes,
like Ptime or Pspace. Different calculi have been used for this
purpose coming from primitive recursion, lambda calculus, rewriting
systems (\textit{e.g.}
\cite{BellantoniCook,MarionMoyen00,LeivantMarion93})\dots{} A
convenient way to see these systems is in general to describe them as
a subset of programs of a larger language satisfying certain criteria:
for instance primitive recursive programs satisfying safe/ramified
recursion conditions, rewriting systems admitting a termination
ordering and quasi interpretation, etc\dots

\textbf{Inference.} To use such ICC systems for programming purpose it is
natural to wish to automatize the verification of the criterion. This way
the user could stick to a simple programming language and the compiler
would check whether the program satisfies the criterion, in which case
a complexity property would be guaranteed.

 In general this decision procedure involves finding a certain
\textit{witness}, like a type, a proof or a termination ordering.
Depending on the system this witness might be useful to provide more
precise information, like an actual bound on the running time, or a
suitable strategy to evaluate the program. It might be used as a
certificate guaranteeing a particular quantitative property of the
program.

\textbf{Light linear logic.} In the present work we consider the
 approach of Light linear logic (\LLL) (\cite{Girard98}), a variant of
 Linear logic which characterises polynomial time computation, within
 the proofs-as-programs correspondence. It includes higher-order and
 polymorphism, and can be extended to a naive set theory
 (\cite{Terui04}), in which the provably total functions correspond to
 the class of polynomial time functions.

The original formulation of \LLL\ by Girard was quite complicated, but a
first simplification was given by Asperti with Light affine logic
(\LAL)  (\cite{AspertiRoversi02}). Both systems have two modalities (one more than Linear logic)
to control duplication. There is a forgetful map to system F terms
(polymorphic types) obtained by erasing some information (modalities)
in types; if an \LAL\ typed term $t$ is mapped to an F-typed term $M$ we
also say that $t$ is a \textit{decoration} of $M$.

So an \LAL\ program can be understood as a system F program, together
with a typing guarantee that it can be evaluated in polynomial time
once that program is written and evaluated in the right syntax (see
below). As system F is a reference system for the study of
polymorphically typed functional languages and has been extensively
studied, this seems to offer a solid basis to \LAL.

 However \LAL\ itself is still difficult to handle and following the
 previous idea for the application of ICC methods, we would prefer to
 use plain lambda calculus as a front-end language, without having to
 worry about the handling of modalities, and instead to delegate the
 \LAL\ typing part to a type inference engine. The study of this
 approach was started in \cite{Baillot02}. For it to be fully
 manageable however several conditions should be fulfilled:

\begin{enumerate}[(1)]
\item  a suitable way to execute the lambda-terms with the expected
complexity bound,
\item  an efficient type inference,
\item a typed language which is expressive enough so that a reasonable 
range of programs is accepted.
\end{enumerate}

 The language \LAL\ presents some drawback for the first point, because the \LAL\ typed
terms need to be evaluated with a specific graph syntax, \textit{proof-nets}, in order
to satisfy the polynomial bound, and plain beta reduction can lead to
exponential blow-up.

In a previous work (\cite{BaillotTerui04}) we
addressed this issue  by defining a subsystem of \LAL, called
Dual light affine logic (\DLAL). It is defined with both linear and
non-linear function types. It is complete for Ptime just as \LAL\ and
its main advantage is that it is also Ptime sound w.r.t. beta
reduction: a \DLAL\ term admits a bound on the length of all its beta
reduction sequences. Hence \DLAL\ stands as a reasonable substitute for
plain \LAL\ for typing issues.

Concerning point 2, as type inference for system F is undecidable 
(\cite{Wells99}) we
do not try to give a fully-fledged type inference algorithm from untyped
terms.  Instead, to separate the polymorphic part issue from the
proper \DLAL\ part one, we assume the initial program to be already typed
in F. Either the system F typing work is left to the user, or one could use a
partial algorithm for system F typing for this preliminary phase.
 
 So the contribution of the present work is to define an efficient
algorithm to decide if a system F term can be decorated in a \DLAL\ typed
term. This was actually one of the original motivations for defining
\DLAL. We show here that decoration can be performed in polynomial
time. This is obtained by taking advantage of intuitions coming from
proof-nets, but it is presented in a standard form with a first phase
consisting in generating constraints expressing typability and a
second phase for constraints solving. 
One difficulty is that the
initial presentation of the constraints involves disjunctions of
linear constraints, for which there is no obvious Ptime bound. Hence
we provide a specific resolution strategy.

The complete algorithm is already implemented in ML, in a way that
follows closely the specification given in the article.  It is modular
and usable with any linear constraints solver.  The code is commented,
and available for public download (Section \ref{l-implement}).  With
this program one might thus write terms in system F and verify if they
are Ptime and obtain a time upper bound. It should in particular be
useful to study further properties of \DLAL\ and to experiment with
reasonable size programs.

 The point 3 stressed previously about expressivity of the system
remains an issue which should be explored further. Indeed the \DLAL\
typing discipline will in particular rule out some nested iterations
which might in fact be harmless for Ptime complexity. This is related
to the line of work on the study of intensional aspects of Implicit
computational complexity (\cite{MarionMoyen00,Hofmann03}). 

 However it might be possible to consider some combination of \DLAL\
with other systems which could allow for more flexibility, and we
think a better understanding of \DLAL\, and in particular of its type
inference, is a necessary step in that direction.

\textbf{Related work.} Inference problems have been studied for
several ICC systems (\textit{e.g.} \cite{Amadio05},
\cite{HofmannJost02}). Elementary linear logic (\EAL,
\cite{Girard98,DanosJoinet99}) in particular is another variant of
Linear logic which characterises Kalmar elementary time and has
applications to optimal reduction. Type inference for propositional
EAL (without second-order) has been studied in
\cite{CoppolaMartini01},\cite{CoppolaRonchi03},\cite{CoppolaDalLagoRonchi05}
and \cite{BaillotTerui05} which gives a polynomial time procedure.
Type inference for \LAL\ was also investigated, in
\cite{Baillot02,Baillot04}.  To our knowledge the present algorithm is
however the first one for dealing with polymorphic types in an
EAL-related system, and also the first one to infer light types in
polynomial time.

 This article is an extended version of the paper
\cite{AtassiBaillotTerui06}.  Its main novelties are the following
ones:
\begin{enumerate}[$\bullet$]
\item it is self-contained and complete proofs are provided for the theorems;
\item a new section has been added discussing the problem of type inference
with data-type domain specification (Section \ref{sect:domainspecification});
\item the section on the implementation of the algorithm and examples of type inference
(Section \ref{l-implement}) has been developed, with in particular more examples like
the predecessor and polynomials.
\end{enumerate}

\textbf{Acknowledgements.} The authors wish to thank Laurent R\'egnier
for useful discussions related to the topic of this paper.

\section{From system F to \DLAL}

 The language $\lf$ of system F types is given by:
$$T, U::= \alpha \; | \; T \rightarrow U \;|\; \forall \al . T \; .$$

We assume that a countable set of term variables
$x^T, y^T, z^T,\ldots$ is given for each type $T$.
The terms of system $F$ are built as follows
(here we write $M^T$ to indicate that the term $M$ has type $T$):
$$
x^T
\quad
(\la x^T. M^U)^{T\rightarrow U}
\quad
((M^{T\rightarrow U}) N^T)^U \quad 
(\La \alpha. M^U)^{\forall \alpha.U}
\quad
((M^{\forall\alpha. U})T)^{U[T/\alpha]},
$$

with the proviso that when building a term
$\La \alpha. M$,
$\alpha$ does not occur free in the types of
free term variables of $M$ (the {\em eigenvariable condition}).
The set of free variables of $M$ is denoted $\FV{M}$.

It is well known that there is no sensible resource bound
(i.e.\ time/space) on the execution of system F terms in general. On
the other hand, we are practically interested in those terms which can
be executed in polynomial time.
  However the class $\mathcal{P}$ of such
terms is neither recursively enumerable nor co-recursively enumerable.
This can be verified for instance in the following way,
by reduction of the problem of solvability of Diophantine equations.
For each Diophantine equation $P(x)=0$, build a lambda term $M_P$ such
 that, when a binary word
$w$ is given,  $M_P(w)$ returns $\epsilon$ if $P(x)=0$ has an integer solution
 $n$ with  $-|w| \leq n \leq |w|$,
and returns a word of length $2^{|w|}$ otherwise. Then $M_P \in \mathcal{P}$ iff $P(x)=0$
 has an integer solution. There is also a complementary reduction,
establishing our claim.
 Actually a stronger result is shown in \cite{BonfanteMarionMoyen}:
the class  $\mathcal{P}$ is    $\Sigma^0_2$-complete.

So  we are
naturally led to the study of sufficiently large subclasses of
$\mathcal{P}$.
  The system \DLAL\ gives such a class in a purely
type-theoretic way.

 The language $\ldlal$ of \DLAL\ types is given by:
$$A, B::= \alpha \; | \; A \fm B \; | \; A \fli B \; |\; \pa A   \;|\; \forall \al . A \;.$$
We note $\pa^0 A=A$ and   $\pa^{k+1} A=\pa \pa^k A$.
 The erasure map $\eras{(.)}$ from $\ldlal$ to $\lf$ is defined by:
$$\eras{(\pa A)}= \eras{A},\ \ \
\eras{(A\fm B)}=\eras{(A\fli B)}=\eras{A} \rightarrow \eras{B},$$
and $\eras{(.)}$ commutes
with the other connectives.
 We say
$A\in \ldlal$ is a {\em decoration} of 
$T\in \lf$ if $\eras{A}=T$.

A {\em declaration} is a pair of the form $x^T: B$ with $B^- = T$. It
is often written as $x: B$ for simplicity.  A {\em judgement} is of
the form $\Gamma;\Delta \vdash M: A$, where $M$ is a system F term,
$A\in \ldlal$ and $\Gamma$ and $\Delta$ are disjoint sets of
declarations. The intuition is that the (free) variables in $\Gamma$ are duplicable
 (non-linear),
 while the ones
in $\Delta$ are not (they are linear).
 When $\Delta$ consists of $x_1:A_1, \ldots, x_n:A_n$,
$\pa\Delta$ denotes $x_1: \pa A_1, \ldots, x_n:\pa A_n$.  The type
assignment rules are given on Figure \ref{NDLALrules}.  Here, we
assume that the substitution $M[N/x]$ used in ($\pa$ e) is {\em
capture-free}. Namely, no free type variable $\alpha$ occurring in $N$
is bound in $M[N/x]$.  We write $\Gamma;\Delta \vdash_{DLAL} M: A$ if
the judgement $\Gamma;\Delta \vdash M: A$ is derivable.

 \begin{figure*}[ht]
\begin{center}
\fbox{
\begin{tabular}{c@{}cc}
  &{\infer[\mbox{(Id)}]{;x^{A^-}:A \vdash x^{A^-}:A}{}\quad} & \\
 &&\\
 &{\infer[\mbox{($\fm$ i)}]{\Gamma;\Delta \vdash \la x^{A^-}.
M: A \fm B }
 {\Gamma;x^{A^-}:A,\Delta \vdash M:B}\quad}
  & {\infer[\mbox{($\fm$ e)}]{\Gamma_1,\Gamma_2;
\Delta_1,\Delta_2\vdash (M) N :B }
  {\Gamma_1;\Delta_1 \vdash M:A \fm B & \Gamma_2;\Delta_2 \vdash N:A}}\\[1ex]
&{\infer[\mbox{($\fli$ i)}]{\Gamma;\Delta \vdash \la x^{A^-}. M: A \fli B }
 {x^{A^-}: A, \Gamma;\Delta  \vdash M:B}\quad}
  & {\infer[\mbox{($\fli$ e) (*)}]{\Gamma, z: C;\Delta \vdash (M) N :B }
  {\Gamma;\Delta \vdash M:A \fli B &  ;z:C \vdash N:A}}\\[1ex]
&{\infer[\mbox{(Weak)}]{\Gamma_1,\Gamma_2;\Delta_1,\Delta_2 \vdash M: A }
 {\Gamma_1;\Delta_1 \vdash M:A & }\quad}
  &{\infer[\mbox{(Cntr)}]{x: A, \Gamma;\Delta \vdash M[x \slash x_1, x \slash x_2] :B }{x_1: A,x_2: A, \Gamma;\Delta \vdash M:B }} \\[1ex]
&{\infer[\mbox{($\pa$ i)}]{\Gamma; \pa\Delta \vdash M: \pa A }
 { ;\Gamma, \Delta \vdash M:A}\quad}
  & {\infer[\mbox{($\pa$ e)}]{\Gamma_1,\Gamma_2;\Delta_1,\Delta_2
     \vdash M[N \slash x] :B }
  {\Gamma_1;\Delta_1 \vdash N: \pa A  & \Gamma_2;x:\pa A,\Delta_2
\vdash M:B}}\\[1ex]
&{\infer[\mbox{($\forall$ i) (**)}]{\Gamma;\Delta \vdash  \La \alpha.M:\forall \alpha. A}{\Gamma;\Delta \vdash M:A}\quad} &
 {\infer[\mbox{($\forall$ e)}]{\Gamma;\Delta \vdash (M)\eras{B} :A[B \slash \al] }
{\Gamma;\Delta \vdash M:\forall \al. A}} \\[1ex]
&\multicolumn{2}{c}{
\mbox{(*) $z:C$ can be absent.}}\\
&\multicolumn{2}{c}{
\mbox{(**) $\alpha$ does not occur free in $\Gamma, \Delta$.}}
 \end{tabular}
}
  \caption{Typing system F terms in \DLAL}\label{NDLALrules}
\end{center}
\end{figure*}

Examples of concrete programs typable in \DLAL\ are given 
in Section \ref{l-implement}.

Recall that binary words, in $\{0,1\}^*$,  can be given in system F the type:
 $$W_F
 = \forall\alpha.
(\alpha\rightarrow\alpha)\rightarrow
(\alpha\rightarrow\alpha)\rightarrow
 (\alpha\rightarrow\alpha) \; .$$
A corresponding type in \DLAL, containing the same terms, is given by:
$$W_{DLAL} = \forall\alpha.
(\alpha\llto\alpha)\Rightarrow
(\alpha\llto\alpha)\Rightarrow
\pa (\alpha\llto\alpha) \; .$$

The {\em depth} $d(A)$ of a \DLAL\ type $A$ is defined by:
\[
\begin{array}{rclrcl}
d(\alpha) & = & 0, & d(\forall\alpha. B) & = & d(B), \\
d(A\fm B) & = & max(d(A),d(B)), & d(\pa A) & = & d(A)+1,\\
d(A\fli B) &=&  max(d(A)+1,d(B)). & & &
\end{array}
\]
A type $A$ is said to be $\Pi_1$ if it does not
contain a negative occurrence of $\forall$; like for instance
$W_{DLAL}$.

The fundamental properties of \DLAL\ are the following
\cite{BaillotTerui04}:
\begin{thm}\label{DLALfundamentalproperties}~
\begin{enumerate}[\em(1)]
\item
 For every function $f:\{0,1\}^* \longrightarrow \{0,1\}^*$
in \textsc{DTIME}$[n^k]$,
there exists a closed term $M$ of type
$W_{DLAL} \llto \S^d W_{DLAL}$ with $d= O(\log k)$ representing $f$.
\item
Let $M$ be a closed term of system F that has a $\Pi_1$ type $A$ in
\DLAL. Then
$M$ can be normalized in $O(|M|^{2^d})$ steps by $\beta$-reduction,
where $d=d(A)$ and $|M|$ is the structural size of $M$.
Moreover, the size of any intermediary term occurring in normalization
is also bounded by $O(|M|^{2^d})$.\qed
\end{enumerate}
\end{thm}

Although \DLAL\ does not capture all Ptime {\em algorithms}
$\mathcal{P}$,
the result 1 guarantees that
 \DLAL\ is at least expressive enough to 
represent all Ptime
{\em functions}.
In fact,
\DLAL\ is as expressive as \LAL\, even
at the level of algorithms, because there exists a generic translation from
\LAL\ to \DLAL\ given by:
$$ (\bs A)^o= \forall \al. ((A^o \fli \al)\fm \al), \qquad (.)^o \mbox{ commutes with other connectives than } \bs. $$
See the full version of \cite{BaillotTerui04} (forthcoming)
for details.

The result 2 on the other hand
implies that if we ignore the embedded types occurring in
 $M$, the normal form of $M$ can be computed in polynomial time
({\em by ordinary $\beta$-reduction}; that is the difference
from \LAL).
 
Now, let $M^{W_F\rightarrow W_F}$ be a system F typed term and suppose
that we know that it has a \DLAL\ type $W_{DLAL} \llto \pa^d W_{DLAL}$
for some $d\geq 0$.  Then, by the consequence of the above theorem, we
know that the term $M$ is Ptime.
In fact, given a binary word $w\in\{0,1\}^*$, 
consider its Church coding 
 $\underline{w}$ of type $W_{DLAL}$. Then 
we have that $(M) \underline{w}$ has type
 $\pa^d W_{DLAL}$, and can thus be evaluated in 
 $O(|\underline{w}|^{2^{d+1}})$ steps.
Thus by assigning a \DLAL\ type to a given system F term, one can
 {\em statically verify} a polynomial time bound for
its execution.

In order to use \DLAL\ for resource
verification of system F terms, we address the following problem:
\begin{prob}[\DLAL\ typing]\label{problemtyping}
Given a closed term $M^T$ of system F, determine if there
is a decoration $A$ of $T$
such that
$\vdash_{DLAL} M:A$.
\end{prob}
(Here the closedness assumption is only for readability.)

In the sequel, we show that there is
a polynomial time algorithm for solving
the \DLAL\ typing problem. 
~\\

\section{Localization of \DLAL\ type inference}

To solve the \DLAL\ typing problem, the main obstacle is 
that the typing rules of \DLAL\ are not syntax-directed.
In particular, the rule ($\pa$ i) does not correspond to 
any constructs of system F terms, and 
the rule ($\pa$ e) involves term substitution.
These features make local reasoning on types impossible.

To overcome the difficulty, we introduce (following
\cite{AspertiRoversi02}) an intermediary syntax which is more
informative than system F terms, but not more informative than \DLAL\
derivations themselves (in \ref{ss-pseudo}).  In particular, it has
explicit constructs for ($\pa$ i).  In addition, we replace the global
typing rules of \DLAL\ (which involve substitution) with some local
typing rules and a set of conditions (in \ref{ss-local} and
\ref{ss-conditions}).  We then show that our Local typing rules and
conditions exactly characterise system F terms typable in \DLAL\ (in
\ref{ss-correctness}).

\subsection{Pseudo-terms}\label{ss-pseudo}

We begin with introducing an intermediary syntax, which consists of 
\DLALS\ types and pseudo-terms.

First we decompose $A\fli B$ into $\bs A\fm B$.
The language $\ldlals$ of \DLALS\ types is given by:
\begin{eqnarray*}
A & ::= & \alpha\ |\ D\fm A\ |\ \forall\alpha.A \ |\ \pa A \; ,\\
D & ::= & A\ |\ \bs A \; .
\end{eqnarray*}
There is a natural map $(.)^\star$
from $\ldlal$ to $\ldlals$
such that $(A\fli B)^\star = \bs A^\star \fm B^\star$
and commutes with the other operations. 
The erasure map $\eras{(.)}$ from $\ldlals$ to $\lf$ can be
 defined as before.
A \DLALS\ type  is called a {\em bang type} if
it is of the form $\bs A$,
and otherwise called a {\em linear type}.
In the sequel, $A,B,C$ stand for linear types, and $D$ for
either bang or linear types.

We assume there is a countable set of term variables
$x^D, y^D, z^D,\dots$ for each $D\in \ldlals$.
The \textit{pseudo-terms} are defined by the following grammar:
$$ t, u::= x^{D} \;|\; \la x^{D}. t \;|\;
(t)u \;|\;
 \La \alpha.t \;|\; (t)A \;|\; \pa t \;|\; \pad t \; ,$$
where $A$ is a linear type and $D$ is an arbitrary one.
 The idea is that
$\pa$
corresponds to the main door of a $\pa$-box (or a $\bs$-box)
in \textit{proof-nets} (\cite{Girard87a,AspertiRoversi02}) while
$\pad$
 corresponds to auxiliary doors. But note
that there is no information in the pseudo-terms to link occurrences
of $\pa$ and $\pad$ corresponding to the same box, nor distinction
between $\pa$-boxes and $\bs$-boxes.

 There is a natural erasure map from pseudo-terms to
system F terms, which we will also denote
by $(.)^-$, consisting in removing all occurrences of $\pa, \pad$,
replacing $x^{D}$ with $x^{D^-}$ and $(t)A$ with $(t)A^-$.
When $t^- = M$, $t$ is called a
\textit{decoration} of $M$.

 Let $t$ be a pseudo-term and $u$ be a subterm of $t$. We say that $u$ is a \textit{door-extreme subterm} 
of $t$ if the following holds: if $u$ is of the form
$u=\pa u'$ or $u=\pad u'$ then $\pa u$ and $\pad u$ are not subterms of $t$.

 As an example consider $t=(x\; \pa \pa y)$. Its door-extreme subterms are 
$\{t, \; x, \; \pa \pa y, \; y \}$, and
$\pa y$ is a subterm of $t$ but not a door-extreme subterm.

For our purpose, it is sufficient to consider
the class of \textit{regular} pseudo-terms, given by:
\begin{eqnarray*}
u & ::= & x^{D}  \;|\;
 \la x^{D}. t \;|\;
(t)t \;|\;
 \La \alpha.t \;|\; (t)A \;, \\
t & ::= & \pa^m u \;, 
\end{eqnarray*}
where $m$ is an arbitrary value in $\ZZ$ and $\pa^m t$ denotes
$\pa\cdots \pa t$ ($m$ times) 
if $m\geq 0$, and 
$\pad\cdots \pad t$ ($-m$ times) if $m<0$.

In other words, a pseudo-term is regular if
and only if it does not contain any subterm of the
form $\pa \pad u$ or $\pad \pa u$.

\subsection{Pseudo-terms and proof-nets}

 In this section we illustrate the links between pseudo-terms and proof-nets.
It is independent  of the sequel and can be skipped without problem.

  The translation $(.)^\star$ from \DLAL\ to \LAL\ gives a mapping on
derivations; therefore a \DLAL\ type derivation corresponds to an
\LAL\ proof and thus to a proof-net (\cite{AspertiRoversi02}).
To facilitate the reading we will use here a `syntax-tree like'
representation for intuitionistic \LAL\ proof-nets.

As an example consider the following term:
$$M= (\la f. (f) ((f) \; x))((\la h. h) \; g) \; . $$
It can be given the typing $x: \pa \al \vdash M: \pa \al $, with the derivation
of Fig.\  \ref{exampleDLALderivation}. The corresponding
(intuitionistic) proof-net is given on Fig.\  \ref{examplePN}. For readers
more familiar with the classical representation of proof-nets (in the style of \textit{e.g.}
\cite{AspertiRoversi02}), the corresponding
representation is given on Fig.\  \ref{exampleCPN}.

\begin{figure}[ht]
{\tiny
\begin{prooftree}
\AxiomC{$;f_2: \al \fm \al \vdash f_2: \al \fm \al$ }
\AxiomC{$;x: \al \vdash x: \al$ }
\BinaryInfC{$; f_2: \al \fm \al , x: \al   \vdash (f_2) \; x: \al  $}
\AxiomC{$;f_1: \al \fm \al \vdash f_1: \al \fm \al$ }
\BinaryInfC{$; f_1: \beta, f_2: \beta, x: \al \vdash (f_1) \; ((f_2) \; x ): \al$}
\UnaryInfC{$ f_1: \beta, f_2: \beta; x: \pa \al \vdash (f_1) \; ((f_2) \; x ): \pa \al$}
\UnaryInfC{$ f: \beta; x: \pa \al \vdash (f) \; ((f) \; x ): \pa \al$}
\UnaryInfC{$ ; x:\pa \al  \vdash\la f.  (f) \; ((f) \; x ):  \beta \fli \pa \al $}

\AxiomC{$;h: \beta\vdash h: \beta$ }
\UnaryInfC{$; \vdash \la h . h:  \beta \fm  \beta $}
\AxiomC{$;g: \beta\vdash g: \beta$ }
\BinaryInfC{$; g: \beta \vdash (\la h . h)\; g: \beta $}
\BinaryInfC{$g: \beta; x: \pa \al \vdash (\la f.  (f) \; ((f) \; x ))  ((\la h . h)\; g): \pa \al $}
\end{prooftree}
}
where $\beta=\al \fm \al .$
\caption{Example: \DLAL\ derivation for $M$.}\label{exampleDLALderivation}
\end{figure}

\begin{figure}[ht]
\begin{center}
\input cpn60.pstex_t
\end{center}
\caption{Classical Proof-Net corresponding to the example.}\label{exampleCPN}
\end{figure}

\begin{figure}[ht]
\begin{center}
\input examplePN75.pstex_t
\end{center}
\caption{Example of Proof-Net.}\label{examplePN}
\end{figure}

 The pseudo-term corresponding to the previous derivation is:
$$ t= (\la f. \pa (\pad f) (( \pad f )\;  \pad x )) \pa ((\la h. h)\;  \pad g) \;.$$
It is represented graphically on Fig.\  \ref{exampledoors}: to $\pa$ and
$\pad$ correspond respectively opening and closing doors.

In a proof-net, a box can be thought of as an opening door connected to a certain number 
(possibly none) of closing doors. If in the proof-net of Fig.\  \ref{examplePN}
we disconnect opening doors from closing doors we get the graph of 
Fig.\  \ref{exampledoors}, corresponding to the pseudo-term.

\begin{figure}[ht]
\begin{center}
\input exampledoorspaths75.pstex_t
\end{center}
\caption{Example: graph corresponding to pseudo-term.}\label{exampledoors}
\end{figure}

Our method for type inference relies on a procedure
for deciding if a pseudo-term comes from a \DLAL\ derivation. This essentially corresponds
to deciding if a pseudo-term corresponds to a proof-net, that is to say
in particular deciding whether opening and closing doors can be matched
in such a way to yield a correct distribution of boxes.
   
\subsection{Local typing condition}\label{ss-local}

We now describe a way to assign types to pseudo-terms 
in a locally compatible way.
A delicate point
in \DLAL\ is that it is sometimes natural to associate {\em two}
types to one variable $x$. For instance, we have
$x:A; \vdash_{DLAL} x:\pa A$ in \DLAL, and this can be read as
$x:\bs A\vdash x:\pa A$ in terms of \DLALS\ types.
We thus distinguish between the 
{\em input types}, which are inherent to variables,
and the {\em output types}, which are
inductively assigned to
all pseudo-terms.
The condition (i) below is concerned with
the output types.
In the sequel,
$D^\circ$ denotes $\pa A$ if
$D$ is of the form $\bs A$,
and otherwise  denotes $D$ itself.

A pseudo-term $t$ satisfies the
{\em Local typing condition}
if the following holds:
\begin{enumerate}[(i)]
\item one can inductively assign a {\em linear} type
to each
subterm of $t$ in the following way (here the notation $t:A$
indicates that $t$ has the output type $A$):
$$
\begin{array}{cccc}
\infer{x^{D}:D^\circ}{}
&
\infer{\la x^{D}. t: D\fm B}{ t: B}
&
\multicolumn{2}{c}{\infer{(t)u:B}{t: D\fm B & u : A & D^\circ = A}}\\[1em]
\infer{\pa t:\pa A}{ t: A}
&
\infer{\pad t:A}{ t: \pa A}
&
\infer{\La \alpha.t: \forall\alpha.A}{t : A}
&
\infer{(t) B: A[B/\alpha]}{t: \forall\alpha.A}
\end{array}
$$
\item when a variable $x$ occurs more than once in $t$,
it is typed as $x^{\bs A}$,
\item $t$ satisfies the eigenvariable condition.
Namely, for any subterm of the form
$\La \alpha. u$ and any free term variable $x^D$ in $u$,
$\alpha$ does not occur free in $D$.
\end{enumerate}
We also say that $t$ is {\em
locally typed}.

The Local typing rules are syntax-directed, and assign a
unique type to each pseudo-term whenever possible.
Notice that 
there is a type mismatch between $D$ and $A$
in the application rule 
when $D$ is a bang type. 
This mismatch will be settled by the \textit{Bang condition}
below. 

\subsection{Boxing conditions}\label{ss-conditions}

It is clear that local typability is not a sufficient condition for 
typability in \DLAL, as it does not ensure that doors $\pa, \pad$
are well placed
so that boxes can be built around them. Moreover, it does not distinguish
$\pa$- and $\bs$- boxes.
We therefore impose additional conditions on locally typed pseudo-terms.
 
We consider words over the language $\mathcal{L}=\{\pa,\pad\}^{*}$ and
$\leq$ the prefix ordering.
If $t$ is a pseudo-term and $u$ is an occurrence of subterm
 in $t$, let $\lis{t}{u}$ be the word inductively defined as follows.
If $t=u$, let $\lis{t}{u} = \epsilon$. Otherwise:
$$\begin{array}{lcl}
\lis{\pa t}{u} &=& \pa ::(\lis{t}{u}), \\
 \lis{\pad t}{u} &=& \pad::(\lis{t}{u}),\\
 \lis{\la y^{D}.t_1}{u} &=&
\lis{\La \alpha.t_1}{u}
 =
\lis{(t_1)A}{u} = \lis{t_1}{u}, \\
\lis{(t_1)t_2}{u} &=& \lis{t_i}{u}, \mbox{ where $t_i$ is the
subterm
containing $u$}.
\end{array}
$$
That is to say, $\lis{t}{u}$ collects the modal symbols
$\pa$, $\pad$ occurring on the path from the root to
the node $u$
in the term tree of $t$.
 We define a map $s: \mathcal{L} \rightarrow \ZZ$ by:
\begin{eqnarray*}
s(\epsilon) &=& 0,\\
s(\pa :: l) &=& 1+ s(l),\\
s(\pad :: l) &=& -1+ s(l).
\end{eqnarray*}
 A word
$l\in\mathcal{L}$ is {\em weakly well-bracketed} if $ \forall l'\leq
l, s(l')\geq 0, $ and is {\em well-bracketed} if this condition holds
and moreover $s(l)=0$: think of $\pa$ and $\pad$ resp.\ as opening and
closing brackets.

\textbf{Bracketing condition.}
Let $t$ be a pseudo-term. We say that $t$ satisfies the
\textit{Bracketing condition} if:
\begin{enumerate}[(i)]
\item for any occurrence of free variable $x$ in $t$,
$\lis{t}{x}$ is well-bracketed;
\item{for any occurrence of
an abstraction subterm $\la x.v$ of $t$:
\begin{itemize}
\item[(ii.a)] $\lis{t}{\la x.v}$ is weakly well-bracketed, and
\item[(ii.b)] for any occurrence of $x$ in $v$,
$\lis{v}{x}$ is well-bracketed.
\end{itemize}}
\end{enumerate}

This condition is sufficient to rule out the canonical morphisms for
dereliction and digging, which are not valid in \DLAL\ (nor in $EAL$):
$$
\la x^{\pa A}. \pad x : \pa A\fm A, \quad\quad
\la x^{\pa A}. \pa x :\pa A\fm \pa\pa A\; .$$
Since $\lis{\pad x}{x}=\pad$ and $\lis{\pa x}{x}=\pa$, they do not
satisfy the Bracketing condition (ii.b).

\begin{rem}\label{rem:pathsingraphs}
 On the graph representation of pseudo-terms, 
conditions (i), (ii.a) and (ii.b) can be 
visualised as conditions of bracketing holding on certain paths of the
graph: for instance condition (ii.b) means that any (top-down) path from
a $\la x$ binder to an edge corresponding to an occurrence of
$x$ is well-bracketed (considering the opening and closing doors). For instance
the pseudo-term graph of Fig.\  \ref{exampledoors} satisfies these conditions; we 
show on the Figure two paths $\gamma_1$, $\gamma_2$ that have to be well-bracketed
according to (ii.b).
\end{rem}

\textbf{Bang condition.} Let $t$ be a locally typed 
pseudo-term.
A subterm $u$ is called a {\em bang subterm}
of $t$ if it occurs as $(t')u$ in $t$
for some $t': \bs A\fm B$.
We say that $t$ satisfies the
\textit{Bang condition} if for any bang subterm $u$ of $t$,
\begin{enumerate}[(i)]
\item  $u$ contains at most one occurrence of free variable $x^{\bs C}$,
and it has a bang type $\bs C$.
\item for any  subterm $v$ of $u$ such that
$v\neq u$ and $v\neq x$,
$s(\lis{u}{v})\geq 1$.
\end{enumerate}

This condition is sufficient to rule out the canonical morphisms for
monoidalness \linebreak $!A\lltensor !B \llto !(A\lltensor B)$ and $\pa A\llto !A$
which are not valid in \LAL\
(the following terms and types are slightly more complicated since $\ldlals$
does not explicitly contain a type of the form $A\llto \bs B$):
$$
\lambda x^{\bs (A\fm B)}.\lambda y^{\bs B\fm C}. \lambda z^{\bs A}.
 (y)\pa((\pad x)\pad z) \; ,
\quad\quad
\lambda x^{\pa A}.\lambda y^{\bs A\fm B}. (y)\pa (\pad x)\; .
$$
In the first pseudo-term,
the bang subterm $\pa((\pad x)\pad z)$ contains more than one free variable.
In the second pseudo-term, the bang subterm $\pa (\pad x)$ has
a free variable $x$ with a linear type.
Hence they both violate the Bang condition (i).

\begin{rem}\label{rem:bangconditionPN}
The intuition behind the Bang condition might be easier to understand
on the graph representation of pseudo-terms. The idea is that
in a proof-net, the argument of a non-linear application
should be enclosed in a box, with at most one free variable,
as in the Example of Fig.\  \ref{examplePN}. This is enforced on the pseudo-term
by Bang conditions (i) and (ii). Condition (ii) indeed forces the root
of the argument of the application to start with an opening door, and 
this opening door can only be matched by a closing door on the edge
 corresponding to the free variable $x$.
\end{rem}

\textbf{$\La$-Scope condition.}
The previous conditions, Bracketing and Bang, would be enough
to deal with boxes in the propositional fragment of \DLAL.
For handling second-order quantification though, we need
a further condition to take into account the sequentiality enforced
by the quantifiers. For instance consider the following two formulas
(the second one is known as {\em Barcan's formula}):
$$
\mbox{(1) }\pa \forall \alpha.  A \fm \forall \alpha. \pa A \; ,
\quad\quad
\mbox{(2) }
\forall \alpha. \pa A \fm \pa \forall \alpha. A \; .
$$
 Assuming $\alpha$ occurs free in $A$, formula (1) is provable while (2)
is not. Observe that we can build the following
 pseudo-terms which are locally typed and
have respectively type (1) and (2):
$$
t_1= \la x^{\pa\forall\alpha.A}.\La \al. \pa((\pad x)\alpha) \; ,\quad\quad
t_2= \la x^{\forall\alpha.\pa A}.\pa \La \al. \pad((x)\alpha) \; .
$$
 Both pseudo-terms satisfy the previous conditions, but
$t_2$ does not correspond to a \DLAL\ derivation.

Let $u$ be a locally typed pseudo-term.
We say that $u$ {\em depends on} $\alpha$ if the type of $u$ contains a free
variable $\alpha$.
We say that a locally typed pseudo-term $t$ satisfies the
\textit{$\La$-scope condition} if: for any subterm $\Lambda \al.u$ of
$t$ and for any subterm $v$ of $u$ that depends on $\al$,
$\lis{u}{v}$ is weakly well-bracketed.

Coming back to our example: $t_1$ satisfies the $\La$-scope condition,
but $t_2$ does not, because $(x)\alpha$ depends on $\alpha$ and nevertheless
$\lis{\pad((x)\alpha)}{(x)\alpha} = \pad$ is
not weakly well-bracketed.

\smallskip
 We now give a reformulation of the Bang condition (ii), which will be useful later:
\begin{lem}\label{lem:bangconditionlemma}
 Assume that $t$ is a locally typed regular pseudo-term that satisfies the Bracketing condition and that $u$
is a bang subterm of $t$ that satisfies the Bang condition (i). If $u$ has a free variable call it $x$. Then
 the Bang condition (ii) holds for $u$ iff:

\textit{ for any door-extreme subterm $v$ of $u$ such that $v\neq u$, $v\neq x$, 
$s(\lis{u}{v})\geq 1$; and $s(\lis{u}{x})=0$, if $u$ has a free variable $x$.}
\end{lem}
\begin{proof}
As for the `only-if' direction, it suffices to show that 
$s(\lis{u}{x})=0$ whenever $u$ has a free variable $x$.
By the Bracketing condition, 
there is a subterm $w$ of $t$ such that
$\lis{w}{x}$ is well-bracketed ($w$ is of the form $\la x. v$, or
 $w=t$ if $x$ is free in $t$). Therefore $s(\lis{w}{u})\geq 0$ 
and $s(\lis{w}{x})=0$, so $s(\lis{u}{x})\leq 0$. Let $u'$ be
the smallest subterm of $u$ strictly containing $x$. We have
$ s(\lis{u}{u'})\geq 1$ and $ -1\leq s(\lis{u'}{x})\leq 1$,
so $s(\lis{u}{x})=0$.

To show the `if' direction, let $v$ be a subterm of $u$ such that $v\neq u$. If $u$
has a free variable $x$  we also assume that $v\neq x$.
If $v$ is a door-extreme subterm then $s(\lis{u}{v})\geq 1$. 
Otherwise there are two door-extreme
subterms $v_1$, $v_2$ of $u$ such that:
\begin{enumerate}[$\bullet$]
\item $v_1 \subseteq v \subseteq v_2$, where $\subseteq$ denotes the
subterm relation,
\item $v_1$ is an immediate distinct door-extreme subterm of $v_2$.
\end{enumerate}
Because of regularity, we have:
\begin{eqnarray*}
\mbox{either } & s(\lis{u}{v_2}) > s(\lis{u}{v}) >s(\lis{u}{v_1}),\\
\mbox{or } & s(\lis{u}{v_2}) < s(\lis{u}{v}) <s(\lis{u}{v_1}).
\end{eqnarray*}
Moreover we know that $s(\lis{u}{v_2})\geq 1$ and $s(\lis{u}{v_1})\geq 0$
(because if $v_1=x$ then $s(\lis{u}{v_1})=0$, and otherwise 
 $s(\lis{u}{v_1})\geq 1$). Therefore we have $s(\lis{u}{v})\geq 1$.
\end{proof}

\subsection{Correctness of the conditions}\label{ss-correctness}

So far we have introduced four conditions on pseudo-terms:
Local typing, Bracketing, Bang and $\La$-scope. 
Let us call a regular pseudo-term satisfying these conditions
{\em well-structured}. It turns out that the well-structured pseudo-terms
exactly correspond to the \DLAL\ typing derivations.

\begin{lem}\label{l-correct}
Let $M_0$ be a system F term. If
$$x_1: A_1,\ldots, x_m : A_m;\;
y_1: B_1,\ldots, y_n : B_n \vdash_{DLAL} M_0: C,$$
then 
there is a decoration $t$ of $M_0$ with type
$C^\star$ and with
free variables $x_1^{\bs A_1^\star},
\ldots, x_m^{\bs A_m^\star}$,
$y_1^{B_1^\star},\ldots, y_n^{B_n^\star}$
which is well-structured.
\end{lem}

\begin{proof}
One can build a (possibly non-regular)
decoration $M^+_0$ of $M_0$
by induction on the derivation.
Depending on the last typing rule used (see Figure \ref{NDLALrules}), 
$M^+_0$ takes one of the following forms:
$$
\begin{array}{llll}
\mbox{(Id)} & x^{A^\star} & \\
\mbox{($\fm$ i)} & \la x^{A^\star}.M^+  &
\mbox{($\fm$ e)} & (M^+)N^+ \\
\mbox{($\fli$ i)} & \la x^{\bs A^\star}.M^+  &
\mbox{($\fli$ e)} & (M^+)\pa N^+[\pad z^{\bs C^\star}/z] \\
\mbox{(Weak)} & M^+  &
\mbox{(Cntr)} & M^+[x/x_1, x/x_2] \\
\mbox{($\forall$ i)} & \La \alpha.M^+  &
\mbox{($\forall$ e)} & (M^+)B^\star \\
\mbox{($\pa$ i)} & 
\pa M^+ [\pad x_i^{\bs A_i^\star}/ x_i, \pad y_j^{\pa B_j^\star}/ y_j]
& \mbox{($\pa$ e)} & M^+[N^+/x],\\
\end{array}
$$
where $M^+$ in ($\pa$ i) has free variables
$x_1^{A_1}, \dots,$ $x_m^{A_m},$ $y_1^{B_1}, \dots,$ $y_n^{B_n}$.

It is easy to verify that 
$M^+_0$ admits Local typing with the output type 
$C^\star$ and has
the 
free variables $x_1^{\bs A_1^\star},
\ldots, x_m^{\bs A_m^\star}$,
$y_1^{B_1^\star},\ldots, y_n^{B_n^\star}$.

Moreover, one can show by induction on the derivation that
$M^+_0$ satisfies the Bracketing, Bang and $\La$-scope conditions.
Let us just remark:
\begin{enumerate}[$\bullet$]
\item The rules ($\fm$ i) and ($\fli$ i) introduce new abstraction terms
$\la x^{A^\star}.M^+$ and $\la x^{\bs A^\star}.M^+$, respectively.
The Bracketing condition (ii.b) for them follows from the Bracketing
condition (i) for $M^+$.
\item The rule ($\fli$ e) introduces a new bang term
$\pa N^+[\pad z^{\bs C^\star}/z]$. It satisfies the Bang condition (i)
because $N$ contains at most one linear variable $z$.
The condition (ii) holds because $N^+$ satisfies the Bracketing condition,
and thus we have $\lis{N^+}{u}\geq 0$ for any subterm occurrence $u$.

Observe also that the Bracketing condition is maintained because the
$\pa$ added before $N^+$ and the $\pad$ added before the variable $z$
match each other, so $z$ remains well-bracketed, and condition (i) is
preserved; since we add a $\pa$ on $N$, condition (ii.a) is maintained
as well; and as bounded variables of $N$ are left unmodified, (ii.b)
is obviously still verified.

We also have to make sure that the substitution of $\pad z$ for $z$ 
does not violate the $\La$-scope condition. It follows from the 
eigenvariable condition for $N$, which ensures that $z$ does not depend
on any bound type variable.

\item The rule (Cntr) conforms to the Local typing condition (ii).
\item The rule ($\forall$ i) introduces a new type abstraction
$\La \alpha.M^+$. The $\La$-scope condition for it follows from
the Bracketing condition for $M^+$.
\item The rule ($\pa$ i) clearly preserves the Bracketing condition.
It is also clear that the substitution involved
 does not cause violation of the Bang condition
(as $x_i$'s and $y_j$'s have linear types in $M^+$, and thus
do not appear in any bang term), and the $\La$-scope condition
(as $x_i$'s and $y_j$'s do not depend on any bound type variable 
due to  the eigenvariable condition).
\item The rule ($\pa$ e) involves substitution.
The term $M^+[N^+/x]$ satisfies  the $\La$-scope condition since
substitution is capture-free, and thus no free type variable 
in $N^+$ becomes bound in $M^+[N^+/x]$.
\end{enumerate}

Finally, the required  regular pseudo-term $t$ is obtained from 
$M^+_0$ by applying the following rewrite rules
as many times as possible:
$$
\pad\pa u\ \longrightarrow\ u, \quad\quad
\pa\pad u\ \longrightarrow u.$$
It is clear that all the conditions are preserved by these
rewritings.
\end{proof}

To show the converse direction, 
the following Lemma plays a crucial role:

\begin{lem}[Boxing]\label{boxinglemma}
If $\pa t: \pa A$ is a well-structured pseudo-term,
then there exist pseudo-terms $v: A$, $u_1: \pa B_1$, \dots, 
$u_n :\pa B_n$,
unique (up to renaming
of $v$'s free variables) such that:
\begin{enumerate}[\em(1)]
\item $FV(v)=\{x_1^{B_1}, \dots, x_{n}^{B_{n}}\}$ and
each $x_i$ occurs exactly once in $v$,
\item $\pa t=
\pa v[\pad u_1/x_1,\dots,\pad u_n/x_n]$
(substitution is assumed to be capture-free),
\item $v, u_1,\dots, u_n$ are well-structured.
\end{enumerate}
\end{lem}

\noindent

\begin{proof}
Given $\pa t$, assign an index to
each occurrence of $\pa$ and $\pad$ in $\pa t$
to distinguish occurrences
(we assume that
the outermost $\pa$ has index $0$).
 By traversing from the root of the syntactic tree, one can 
find closing brackets $\pad_1, \dots, \pad_{n}$
that match the opening
bracket $\pa_0$ in $\pa_0 t$.
Replace each $\pad_i u_i :B_i$ with
a fresh and distinct free variable $x_i^{B_i}$ ($1\leq i\leq n$),
and let $\pa v$ be the resulting pseudo-term.
This way one can obtain
$v$, $u_1$, \dots, $u_n$,
 such that condition (2) holds.

Strictly speaking, it has to be checked that the substitution does not
cause capture of type or term variables.  Let us consider the case of
type variables: suppose that $u_i$ contains a subterm $s$ that depends
on a bound variable $\alpha$ of $\pa v$. Then $\pa_0 t$ contains a
subterm of the form $\La \al.v'[\pad_i u_i[s]/x_i]$.  However,
$\lis{v''}{s}$ with $v'' = v'[\pad_i u_i[s]/x_i]$ cannot be weakly
well-bracketed because $\pad_i$ has to match the outermost opening
bracket $\pa_0$. This contradicts the $\La$-scope condition for $\pa_0
t$. Hence the case of type variable capture is solved. A similar argument
using the Bracketing condition shows that the substitutions do no cause
term variable capture either.

As to condition (1), we claim that $v$ does not contain a free variable
other than $x_1, \dots, x_n$.  If there is any, say $y$, then it is
also a free variable of $t$, thus the Bracketing condition for $\pa_0
t$ implies that $\lis{\pa_0 t}{y}$ is well-bracketed, and thus there
is a closing bracket that matches $\pa_0$ in the path from $\pa_0 t$
to $y$.  That means that $y$ belongs to one of $u_1$, \dots, $u_n$,
not to $v$. A contradiction.

Let us now check condition (3). 
As to the Bracketing condition (i) for $v$, let $l_i = 
\lis{\pa_0 t}{\pad_i u_i}$ for each $1\leq i \leq n$.
Then we have $s(l)\geq 1$ for all $\epsilon \neq l\leq l_i$ and 
$s(l_i)=1$, and the same is true of the list $\lis{\pa_0 v}{x_i}$.
Therefore, $\lis{v}{x_i}$ is well-bracketed for each $1\leq i \leq n$.
(ii.a) and (ii.b) are easy.
As for $u_i$ ($1\leq i \leq n$), notice that
$s(\lis{\pa_0 t}{u_i}) = 0$. This means that
for any subterm occurrence $u'$ of $u_i$, we have
$s(\lis{u_i}{u'}) = s(\lis{\pa_0 t}{u'})$. Therefore, 
the Bracketing condition for $u_i$ reduces to that for $\pa_0 t$.

The $\La$-scope condition for $v, u_1, \dots, u_n$ easily reduces to
that for $\pa_0 t$.

As to the Local typing condition, the only nontrivial point to
check is  whether $v$ satisfies the eigenvariable condition.
Suppose that $x_i$ depends on a variable $\al$
which is bound in $v$.
Then $\pa_0 t$ contains a subterm
of the form $\La \al.v'[\pad_i u_i/x_i]$ and $u_i$ depends on $\alpha$.
However, $\lis{v''}{u_i}$
with $v'' = v'[\pad_i u_i/x_i]$ cannot be weakly well-bracketed
because $\pad_i$ should match the outermost opening bracket
$\pa_0$. This contradicts the $\La$-scope condition
for $\pa_0 t$.

To show the Bang condition for $v$ (it is clear for $u_1, \dots,
u_n$), suppose that $v$ contains a bang subterm $v'$.  We claim that
$v'$ does not contain variables $x_1, \dots, x_n$.  If it contains
any, say $x_i$, then $\pa_0 t$ contains $v''=v'[\pad_i u_i/x_i]$ and the
Bang condition for $\pa_0 t$ implies that $s(\lis{v''}{\pad_i u_i})\geq
1$. On the other hand, we clearly have
$s(\lis{\pa_0 t}{v''})\geq 1$ because $v''$ contains the closing
bracket $\pad_i$ that matches $\pa_0$.  As a consequence, we have
$s(\lis{\pa_0 t}{\pad_i u_i})\geq 2$.  This means that $\pad_i$ does not
match $\pa_0$, a contradiction.  As a consequence, $v'$ does not
contain $x_1, \dots, x_n$. So $v'$ occurs in $\pa_0 t$, and
therefore satisfies the Bang condition.
\end{proof}

Now we can prove:

\begin{thm}\label{t-correct}
Let $M$ be a system F term. Then
$$x_1: A_1,\ldots, x_m : A_m;\;
y_1: B_1,\ldots, y_n : B_n \vdash_{DLAL} M: C$$
 if and only if
there is a decoration $t$ of $M$ with type
$C^\star$ and with
free variables $x_1^{\bs A_1^\star},
\ldots, x_m^{\bs A_m^\star}$,
$y_1^{B_1^\star},\ldots, y_n^{B_n^\star}$
which is well-structured.
\end{thm}

\begin{proof}
The `only-if' direction has already been proved.
As for the `if' direction, we prove the following:
if a pseudo-term $t: C^\star $ is well-structured and
$FV(t)=\{x_1^{!A_1^\star },
\dots, x_m^{!A_m^\star }, y_1^{B_1^\star },\dots, \linebreak y_n^{B_n^\star }\}$
for some \DLAL\ types $A_1, \dots, A_m, B_1, \dots, B_n$, 
then we have
$\Gamma;\Delta \vdash_{DLAL} t^-: C$,
where $\Gamma = x_1 : A_1, \dots, x_m : A_m$ and
$\Delta = y_1 : B_1, \dots, y_n : B_n$.
The proof proceeds by induction on the size of $t$.
\begin{enumerate}[$\bullet$]
\item When $t = x_i^{\bs A_i^\star}$ for some $1\leq i\leq m$, 
$C^\star $ must be $\pa A_i^\star $ by Local typing,
and we have $\Gamma;\Delta \vdash_{DLAL} x_i: \pa A_i$.
Likewise, if $t= y_j^{B_j^\star}$ for some $1\leq j\leq n$, 
we have $\Gamma;\Delta \vdash_{DLAL} y_j: B_j$.

\item When $t = \la z^{!A_0^\star }.u : !A_0^\star  \fm C_0^\star$, 
$u : C_0^\star$ is also well-structured; 
observe in particular that the Bracketing condition for $t$ implies
the same for $u$.
By induction hypothesis, we have
$z: A_0,\Gamma; \Delta\vdash_{DLAL} u^-: C_0$, and hence 
$$\Gamma; \Delta\vdash_{DLAL} \la z^{A_0^-}. u^-: A_0\Rightarrow C_0.$$
The case when $z$ has a linear type is similar.

\item When $t = \Lambda \alpha.u : \forall\alpha. C_0^\star$,
$u: C_0^\star$ is also well-structured. Hence one can argue as above;
notice in particular that
the eigenvariable condition on $t$ ensures that one can
apply the rule ($\forall$ i) to $u^-$.

\item When $t = (u) B^\star : C_0^\star [B^\star/\alpha]$, 
$u: \forall\alpha. C_0^\star$ is well-structured,
and the induction hypothesis
yields 
$\Gamma; \Delta \vdash_{DLAL} u^-: \forall\alpha. C_0$.
We therefore obtain 
$\Gamma; \Delta \vdash_{DLAL} (u^-) B^-:  C_0[B/\alpha]$.

\item It is impossible to have $t= \pad u$, because it clearly violates
the Bracketing condition.

\item When $t= \pa t' :\pa C_0^\star$, the Boxing Lemma gives us well-structured
terms $v: C_0^\star$, $u_1:\pa C_1^\star $, \dots, $u_k: \pa C_k^\star $
such that
\begin{enumerate}[(1)]
\item $FV(v)=\{z_1^{C_1^\star }, \dots, z_{k}^{C_{k}^\star }\}$ and
each $z_i$ occurs exactly once in $v$,
\item $\pa t'=
\pa v[\pad u_1/z_1,\dots,\pad u_k/z_k]$.
\end{enumerate}
By the induction hypothesis, we have
$$ ; z_1 : C_1, \dots, z_k : C_k \vdash_{DLAL} v^-: C_0
\quad\quad\mbox{and}\quad\quad
\Gamma; \Delta_i \vdash_{DLAL} u_i^-: \pa C_i$$
for $1\leq i\leq k$, where $(\Delta_1, \dots, \Delta_k)$ is a 
partition of $\Delta$ such that each $\Delta_i$ contains the 
free variables occurring in $u_i$.
Hence by rules ($\pa$ i), ($\pa$ e) and (Cntr), we obtain
$$
\Gamma; \Delta \vdash_{DLAL} v^-[u_1^-/z_1, \dots, u_k^-/z_k] : \pa C_0.
$$

\item 
When $t = (t') t''$ and $t''$ is not a bang subterm,
one can argue as above. 
When $t''$ is a bang subterm, $t'$ and $t''$ are locally typed as
$t':!A^\star \llto C^\star $
and $t'': \pa A^\star $. They are well-structured, and moreover:
\begin{enumerate}[(i)]
\item  $t''$ contains at most one free variable $x_i^{\bs A_i^\star }$,
which is among $\{x_1, \dots, x_m\}$,
\item for any  subterm $v$ of $t''$ such that
$v\neq t''$ and $v\neq x_i$,
$s(\lis{u}{v})\geq 1$.
\end{enumerate}
By the induction hypothesis on $t'$ (and by the fact that $t''$
does not contain any variable of linear type), we have
$$\Gamma;\Delta\vdash_{DLAL} (t')^-: A\Rightarrow C.$$
On the other hand, the 
condition (ii) above entails that
$t''$ is either the variable $x_i$ or of the form $\pa u$.
In the former case, $A^\star  = A_i^\star $ and we have:
$$
\infer{\Gamma; \Delta \vdash (t')^- x_i : C}{
\Gamma;\Delta\vdash (t')^-: A\Rightarrow C
& 
;x_i:A\vdash x_i:A}.
$$
In the latter case, we can apply the Boxing Lemma.
Then the conditions (i) and (ii) entail that 
there is a well-structured term $v: A^\star$ 
with a free variable $z$ such that
$t'' = \pa u = \pa v[\pad x_i/z]$.
Notice here that $z$ has a linear type $A_i^\star$, and 
by renaming, one can assume w.l.o.g. that $z=x_i$ in $v$.
Therefore, we obtain:
$$
\infer{\Gamma; \Delta \vdash (t')^- v^- : C}{
\Gamma;\Delta\vdash (t')^-: A\Rightarrow C
& 
;x_i:A_i\vdash v^-:A}.
$$
\end{enumerate}
\end{proof}

As a consequence of Theorem \ref{t-correct}, 
our \DLAL\ typing problem (Problem \ref{problemtyping}) boils down to:

\begin{prob}[decoration]\label{decorationproblem}
Given a system F term $M$, determine if there exists
a decoration $t$ of $M$ which
is well-structured.
\end{prob}

\section{Parameterization and constraints generation}

To solve the decoration problem (Problem \ref{decorationproblem}), 
one needs to explore an infinite set of decorations.
This can be effectively done 
by introducing an abstract kind of types and terms 
with symbolic parameters (in \ref{ss-parameter}), 
and expressing the conditions for
such abstract terms to be materialized by boolean and
integer constraints over those parameters (in \ref{ss-condition1} and
in \ref{ss-condition2}).

\subsection{Parameterized terms and instantiations}\label{ss-parameter}

Let us begin with introducing a term syntax with parameters.
We use two sorts of parameters: {\em integer parameters} $\prm{n}, \prm{m},\dots$
meant to range over $\ZZ$,
and {\em boolean parameters} $\prm{b_1}, \prm{b_2},\dots$ meant to range over
$\{0,1\}$. We also use
{\em linear combinations of integer parameters}
$\prm{c} = \prm{n_1} + \cdots + \prm{n_k}$, where $k\geq 0$ and each $\prm{n_i}$ is an
integer parameter. In case $k=0$, it is written as $\prm{0}$.

The set of {\em parameterized types} ({\em p-types} for short)
is defined by:
\begin{eqnarray*}
F & :: =&  \alpha\ |\
D\fm A \ |\
\forall\alpha. A \; ,\\
A & :: = & \pa^{\prm{c}} F \; , \\
D & :: = & \pa^{\prm{b},\prm{c}} F \; .
\end{eqnarray*}
 where
$\prm{b}$ is a boolean parameter and
$\prm{c}$ is a linear combination of integer parameters.
Informally speaking,  the parameter $\prm{c}$ in $\pa^{\prm{b},\prm{c}} F$
 stands for the number of modalities ahead of the type, while the boolean
parameter $\prm{b}$ serves to determine whether the first modality, if any, is
 $\pa$ or $\bs$. 
In the sequel, $A, B, C$ stand for {\em linear p-types}
of the form $\pa^{\prm{c}} F$, and $D$ for {\em bang p-types}
of the form $\pa^{\prm{b},\prm{c}} F$, and $E$ for arbitrary p-types.

When $A$ is a linear p-type $\pa^{\prm{c}} F$,
$B[A/\alpha]$ denotes a p-type obtained by
replacing each $\pa^{\prm{c'}}\alpha$ in $B$ with
$\pa^{\prm{c'}+\prm{c}}F$ and
each $\pa^{\prm{b},\prm{c'}}\alpha$ with
$\pa^{\prm{b},\prm{c'}+\prm{c}}F$.
When $D=\pa^{\prm{b},\prm{c}} F$, $D^\circ$
denotes the linear p-type $\pa^{\prm{c}} F$.

We assume that
there is a countable set of variables $x^{D},
y^{D}, \dots$
for each bang p-type $D$.
The \textit{parameterized pseudo-terms} ({\em p-terms} for short) $t, u\dots$
are defined by the following grammar:
\begin{eqnarray*}
u & ::= & x^{D} \;|\; \la x^{D}. t \;|\;
(t) t \;|\;
 \La \alpha.t \;|\; (t)A \;, \\
t & ::= & \pa^\prm{m} u \;.
\end{eqnarray*}

We denote by $\bparam{t}$
the set of boolean
parameters of $t$, and
by $\inparam{t}$ the set of integer parameters of $t$.
An {\em instantiation}
$\phi=(\phi^{b}, \phi^i)$ for a p-term $t$ is given
by two maps $\phi^b:\bparam{t}\rightarrow \{0,1\}$
 and $\phi^i:\inparam{t}\rightarrow \ZZ$.
The map $\phi^i$ can be naturally extended to linear combinations
$\prm{c} = \prm{n_1} + \cdots + \prm{n_k}$ by $\phi^i(\prm{c}) =
\phi^i(\prm{n_1}) + \cdots + \phi^i(\prm{n_k})$.  An instantiation
$\phi$ is said to be {\em admissible} for a p-type $E$ if for any
linear combination $\prm{c}$ occurring in $E$, we have
$\phi^i(\prm{c})\geq 0$, and moreover whenever $\pa^{\prm{b},\prm{c}}
F$ occurs in $E$, $\phi^b(\prm{b})=1$ implies $\phi^i(\prm{c})\geq 1$.
When $\phi$ is admissible for $E$, a type $\phi(E)$ of \DLALS\ is
obtained as follows:
$$
\begin{array}{rclrcll}
\phi(\pa^{\prm{c}} F) & = & \pa^{\phi^i(\prm{c})} \phi(F), \qquad & \phi(\pa^{\prm{b},\prm{c}} F) & = & 
\pa^{\phi^i(\prm{c})} \phi(F) & \mbox{\ \ \ if $\phi^b(\prm{b})=0$,} \\
&&& & = & 
\bs\pa^{\phi^i(\prm{c})-1} \phi(F) & \mbox{\ \ \ otherwise,}
\end{array}
$$
and $\phi$ commutes with the other connectives.
An instantiation $\phi$ for a p-term $t$ is said to be {\em
admissible} for $t$ if it is admissible for all p-types occurring in
$t$.  When $\phi$ is admissible for $t$, a regular pseudo-term
$\phi(t)$ can be obtained by replacing each $\pa^{\prm{m}} u$ with
$\pa^{\phi^i(\prm{m})} u$, each $x^D$ with $x^{\phi(D)}$, and each
$(t)A$ with $(t)\phi(A)$.

 As for pseudo-terms there is an erasure map $(.)^-$ from p-terms
with their p-types to system F terms consisting in forgetting modalities
and parameters.

A {\em free linear decoration} ({\em free bang decoration}, resp.)
of a system F type $T$ is a linear p-type (bang p-type, resp.)
$E$ such that (i) $E^- = T$,
(ii) each linear combination $\prm{c}$ occurring in $E$
is a single integer parameter $\prm{m}$,
and (iii) the parameters occurring in $E$ are mutually distinct.
Two free decorations $\lift{T}_1$ and $\lift{T}_2$
are said to be {\em disjoint}
if the set of parameters occurring in $\lift{T}_1$ is
disjoint from the set of parameters in $\lift{T}_2$.

The {\em free decoration} $\lift{M}$ of a system F term $M$
(which is unique up to renaming of parameters) is obtained as follows:
first, to each variable $x^T$ we associate a parameterized variable
 $\lift{x^T} = x^D$ in such
a way that (i) $D$ is a free bang decoration of $T$, and
(ii) whenever $x^{T_1}$ and $y^{T_2}$ are distinct variables,
the free bang decorations  $D_1$, $D_2$ associated to them are disjoint.
$\lift{M}$ is now defined by induction on the construction of $M$:
\[
\begin{array}{rclrcl}
\lift{\la x^T.M} & = & \pa^\prm{m}
\la x^{\lift{T}}. \lift{M}, \quad  &
\lift{(M)N} & = & \pa^\prm{m} ((\lift{M})\lift{N}), \\
\lift{\La \alpha.M} & = & \pa^\prm{m} \La\alpha.\lift{M}, \quad  &
\lift{(M)T} & = & \pa^\prm{m}((\lift{M})A),
\end{array}
\]
where all newly introduced parameters $\prm{m}$
are chosen to be fresh, and the p-type $A$ in the definition 
of $\lift{(M)T}$ is a free linear decoration of $T$ which is 
disjoint from all p-types appearing in $\lift{M}$.

The key property of free decorations is the following:
\begin{thm}\label{instantiationtheorem}
Let $M$ be a system F term and $t$ be a regular pseudo-term.
Then $t$ is a decoration
of $M$ if and only if
there is an admissible instantiation $\phi$
for $\lift{M}$ such that $\phi(\lift{M})= t$.
\end{thm}

\begin{proof}
We first prove that for any system F type $T$,
any free bang decoration $D$ of $T$ 
and any \DLALS\ type $E$, we have $E^- =T$ iff
there is an admissible instantiation $\phi$ for $D$ such that
$\phi(D) = E$. 
This statement,
as well as a similar one with respect to free linear decorations
and linear \DLALS\ types,
can be simultaneously proved by induction on $T$.
Then the Theorem can be shown by induction on $M$.
\end{proof}

Hence our decoration problem (Problem \ref{decorationproblem}) 
boils down to:
\begin{prob}[instantiation]\label{instantiationproblem}
Given a system F term $M$, determine if there exists
an admissible instantiation $\phi$ for $\lift{M}$ such that
$\phi(\lift{M})$ is well-structured.
\end{prob}

 For that we
will need to be able to state the four conditions (Local typing,
Bracketing, Bang, and $\Lambda$-scope)
on p-terms;
they will yield some constraints on parameters. In the sequel, 
we will speak
of \textit{linear inequations}, meaning in fact both linear equations
and linear inequations.

\subsection{Local typing constraints}\label{ss-condition1}
First of all, we need to express the unifiability of two p-types $E_1$
and $E_2$. 
We define a set
$\unif{E_1}{E_2}$ of constraints by
\begin{eqnarray*}                               
\unif{\alpha}{\alpha} &=& \emptyset,\\
\unif{D_1 \fm A_1}{D_2 \fm A_2}
&=& \unif{D_1}{D_2} \cup \unif{A_1}{A_2},\\
\unif{\forall \alpha.A_1}{\forall \alpha.A_2} &=&\unif{A_1}{A_2},\\
\unif{\pa^\prm{c_1} F_1}{\pa^\prm{c_2} F_2} &=&
\{\prm{c_1}=\prm{c_2}\} \cup
 \unif{F_1}{F_2},\\
\unif{\pa^{\prm{b_1},\prm{c_1}} F_1}{\pa^{\prm{b_2},\prm{c_2}} F_2} &=&
\{\prm{b_1}=\prm{b_2},\prm{c_1}=\prm{c_2}\} \cup\unif{F_1}{F_2}.
\end{eqnarray*}
It is undefined otherwise.
It is straightforward to observe:
\begin{lem}
 Let $E_1$, $E_2$ be two linear (bang, resp.) p-types 
such that $E_1^- = E_2^-$.
Then 
$\unif{E_1}{E_2}$ is defined. Moreover,
when $\phi$ is an admissible
instantiation for $E_1$ and $E_2$,
we have $\phi(E_1)=\phi(E_2)$ if and only if
$\phi$ is a
solution of $\unif{E_1}{E_2}$.
\end{lem}

\begin{proof} By induction on $E_1$.
\end{proof}

For any p-type $E$, define
$$\Adm(E) =
 \{\prm{c}\geq \prm{0} : \mbox{ $\prm{c}$ occurs in
$E$}\}\cup
 \{\prm{b}=\prm{1}\Rightarrow\prm{c}\geq \prm{1} : \mbox{
$\pa^{\prm{b},\prm{c}} F$ occurs in $E$}\}.$$
Then $\phi$ is admissible for $E$ if and only if
$\phi$ is a solution of $\Adm(E)$.

Now consider the free decoration $\lift{M}$ of
a system F typed term $M$.
We assign
to each subterm $t$
of $\lift{M}$ a {\em linear} p-type $B$
and a set $\mc$ of constraints
(indicated as $t:B : \mc$)
as on Figure \ref{localtypingconstraints}. 
Notice that any linear p-type is of the form
$\pa^{\prm{c}} F$. Moreover, since $t$ comes from 
a system F typed term, we know that $t$ has a p-type 
$\pa^{\prm{c}}(D\fm B)$ when 
$t$ occurs as $(t)u$, and
$\pa^{\prm{c}}(\forall\alpha. B)$ when 
$t$ occurs as $(t)A$. In the former case,
we have $(D^\circ)^- = A^-$, so that 
$\unif{D^\circ}{A}$ used in the application rule
is always defined. As a consequence, for any $M$
a unique p-type and 
a unique set of constraints $\mc(\lift{M})$ are obtained.
Finally, observe that
$\lift{M}$ satisfies the eigenvariable condition.

\begin{figure*}
\vspace{-5mm}
$$
\begin{array}{cc}
\infer{x^D : D^\circ : \Adm(D)}{} & 
\infer{\pa^\prm{m} t : \pa^{\prm{m}+\prm{c}}F : \mc\cup\{\prm{m}+\prm{c}\geq 0\}}{
t: \pa^\prm{c}F : \mc} \\[1em]
\infer{\la x^{D}. t: \pa^{\prm{0}} (D \fm A) : \mc\cup \Adm(D)}{ 
t: A : \mc} &
\infer{(t)u : B : \mc_1\cup\mc_2\cup\{\prm{c}=\prm{0}\}\cup \unif{D^\circ}{A}}{
t : \pa^{\prm{c}} (D\fm B):\mc_1 & u : A : \mc_2} \\[1em]
\infer{\La \alpha.t : \pa^{\prm{0}}\forall\alpha. A : \mc}{
 t: A : \mc} &
\infer{(t)A : B[A/\alpha]: \mc\cup\{\prm{c}=\prm{0}\}\cup\Adm(A)}{
  t: \pa^{\prm{c}}(\forall\alpha. B) : \mc}
\end{array}
$$
\vspace{-3mm}
 \caption{Local typing and $\mc(t)$ constraints.}\label{localtypingconstraints}
\end{figure*}

Let $\Ltype(\lift{M})$ be 
$\mc(\lift{M})\cup\{ \prm{b}=\prm{1} : $
$x^{\pa^{\prm{b},\prm{c}}F}$ occurs more than once in $\lift{M}\}$.

\begin{lem}
Let $M$ be a system F term and
$\phi$ be an
instantiation for $\lift{M}$.
Then $\phi$ is admissible for $\lift{M}$ and
$\phi(\lift{M})$ satisfies the Local typing condition
if and only if
$\phi$ is a solution of $\Ltype(\lift{M})$.
\end{lem}

\subsection{Boxing constraints}\label{ss-condition2}
We consider the words over integer parameters $\prm{m}$, $\prm{n}$
\dots, whose set we denote by $\mathcal{L}_p$.
Let $t$ be a p-term
and
$u$ an occurrence of subterm of $t$.
We define, as for pseudo-terms, the word
 $\lis{t}{u}$ in
$\mathcal{L}_p$ as follows. 
If $t=u$, let $\lis{t}{u} = \epsilon$. Otherwise:
$$\begin{array}{lcl}
\lis{\pa^\prm{m} t}{u} &=& \prm{m}::(\lis{t}{u}), \\
 \lis{\la y^{D}.t_1}{u} &=& 
\lis{\La \alpha.t_1}{u} = \lis{(t_1)A}{u} = \lis{t_1}{u}, \\
\lis{(t_1)t_2}{u} &=& \lis{t_i}{u} \mbox{ where $t_i$ is the
subterm containing $u$}. 
\end{array}
$$
The sum $s(l)$ of an element $l$ of $\mathcal{L}_p$ is a linear combination
of integer parameters
defined by: 
$$s(\epsilon) = \prm{0}, \quad\quad s(\prm{m} :: l) = \prm{m}+ s(l).$$
For each  list $l\in \mathcal{L}_p$,
define $\wbracket (l) = \{ s(l')\geq \prm{0}\ |\ l'\leq l\}$
and $\bracket (l) = \wbracket (l) \cup \{s(l)=\prm{0}\}$.

Given a system F term $M$,
we consider the
following sets of constraints:\\

\textbf{Bracketing constraints}.
$\Bra(\lift{M})$ is the union of the following sets:
\begin{enumerate}[(i)]
\item for each occurrence of free variable $x$ in $\lift{M}$,
$\bracket(\lis{\lift{M}}{x})$;
\item[(ii)]{for each occurrence of
an abstraction subterm $\la x.v$ of $\lift{M}$:
\begin{itemize}
\item[(ii.a)] $\wbracket(\lis{\lift{M}}{\la x.v})$,
\item[(ii.b)] for any occurrence of $x$ in $v$, $\bracket(\lis{v}{x})$.
\end{itemize}}
\end{enumerate}

\textbf{Bang constraints}.
A subterm $u$ that occurs in $\lift{M}$ as $(t)u$ with 
$t: \pa^{\prm{c'}}(\pa^{\prm{b},\prm{c}} F\fm B)$ 
is called a {\em bang subterm} of $\lift{M}$ with the {\em critical
parameter} $\prm{b}$. Now
$\Bang(\lift{M})$ is the union of the following sets:
for each bang subterm
$u$ of $\lift{M}$ with a critical parameter $\prm{b}$,
\begin{enumerate}[(i)]
\item $\{\prm{b}=\prm{0}\}$ if $u$ has strictly more than
one occurrence of free variable, and\\
$\{\prm{b}=\prm{1} \Rightarrow
\prm{b'}=\prm{1}\}$ if $u$ has exactly one occurrence of free
variable $x^{\pa^{\prm{b'},\prm{c'}} F'}$.
\item
$\{\prm{b}=\prm{1} \Rightarrow s(\lis{u}{v})\geq \prm{1}\ :\ $
$v$  subterm of $u$ such that $v\neq u$ and $v\neq x\} \cup$

$\{\prm{b}=\prm{1} \Rightarrow s(\lis{u}{x})=\prm{0}\}$.

(where $x$ is the free occurrence of variable in $u$, if there
 is one, otherwise the second set in the union is removed)

\end{enumerate}

\begin{rem}\label{rem:subtermspterms}
Note that if $t$ is a p-term and $\phi$ is an instantiation, the 
pseudo-term $\phi(t)$ might have more subterms than $t$. In fact
subterms of the p-term $t$ are in correspondence with
\textit{door-extreme subterms} of the regular pseudo-term $\phi(t)$.

For instance if $t=\pa^{\prm{m}} x$ and $\phi(\prm{m})=2$,
the subterms of $t$ and $\phi(t)$ are respectively 
$\{\pa^{\prm{m}}x, x\}$ and $\{\pa^{2}x, \pa x, x\}$. The door-extreme
subterms of $\phi(t)$ are $\{\pa^{2}x, x\}$.

 This is why we had to add in the Bang constraints (ii) the condition 
$\{\prm{b}=\prm{1} \Rightarrow s(\lis{u}{x})=\prm{0}\}$ (see Lemma 
\ref{lem:bangconditionlemma}).
\end{rem}

\textbf{$\La$-Scope constraints}.  
$\Scope (\lift{M})$ is the union of the following sets:
\begin{enumerate}[(i)]
\item $\wbracket (\lis{u}{v})$
for each subterm $\La \alpha.u$ of $\lift{M}$ and
 for each subterm $v$ of $u$ that depends on $\alpha$.
\end{enumerate}

We denote
 $\Const(\lift{M})=\Ltype(\lift{M})\cup
\Bra (\lift{M})\cup \Bang (\lift{M}) \cup \Scope (\lift{M})$. 

\begin{rem}\label{rem:pathsinparameterizedgraphs}
 Instead of using p-terms, the parameters and constraints might be visualized
on pseudo-terms graphs. Using our running example of Fig.\  \ref{exampledoors}
we can decorate it with parameters $\prm{m_i}$: see 
Fig.\  \ref{exampleparam}. Each $\prm{m_i}$ stands for
a possible sequence of doors: if it is instantiated with $k\geq 0$ (resp. 
$k\leq -1$) then this will correspond to $k$ (resp. $(-k)$) opening (resp.
closing) doors. Then, for instance, the  Bracketing constraints correspond to
conditions on the parameters occurring along certain paths of the graph
(as indicated in Remark \ref{rem:pathsingraphs}). As an example
the Bracketing constraint (ii.b) for the binder $\la f$ and the two
free occurrences of $f$ gives here
$\{\prm{m_3}\geq 0, \; \prm{m_3}+ \prm{m_4}= 0, \; \prm{m_3}+ \prm{m_5}\geq 0,
\;  \prm{m_3}+ \prm{m_5}+\prm{m_6}= 0\}$.
\end{rem}

\begin{figure}[ht]
\begin{center}
\input exampleparam50.pstex_t
\end{center}
\caption{Example of parameterized pseudo-term graph.}\label{exampleparam}
\end{figure}

\begin{thm}\label{t-param}
Let $M$ be a system F term and $\phi$ be an instantiation for
$\lift{M}$.
Then: $\phi$ is admissible for $M$ and
$\phi(\lift{M})$ is well-structured
if and only if
$\phi$ is a solution of $\Const(\lift{M})$.

Moreover,
the number of (in)equations in $\Const(\lift{M})$ is quadratic
in the size of $M$.
\end{thm}

\begin{proof} Clearly,
the above constraints are 
almost direct parameterizations of
the corresponding conditions given in the previous section.
Let us just examine the Bang condition.

Suppose that $\phi(\lift{M})$ satisfies the Bang condition.
For each (parameterized) bang subterm $u$ 
with the critical parameter $\prm{b}$ in $\lift{M}$,
one of the following two happens:
\begin{enumerate}[$\bullet$]
\item $\phi(u)$ is not a bang subterm of $\phi(\lift{M})$
(in the sense of the previous section). Namely, 
$\phi(\prm{b})= 0$. In this case, $\phi$ is a solution of the equation
$\prm{b}=\prm{0}$, and also of
$\prm{b}=\prm{1} \Rightarrow
\prm{b'}=\prm{1}$ if $u$ has a free
variable $x^{\pa^{\prm{b'},\prm{c'}} F'}$.
\item $\phi(u)$ is a bang subterm of $\phi(\lift{M})$.
Namely, $\phi(\prm{b})= 1$. 
In this case, $u$ contains at most one variable by the Bang condition.
Hence the equation $\prm{b}=\prm{0}$ does not belong to 
$\Bang (\lift{M})$. Moreover, if
$u$ has a free
variable $x^{\pa^{\prm{b'},\prm{c'}} F'}$, $\phi(x)$ must have
a bang type and so
$\phi(\prm{b'})=1$. Hence $\phi$ is a solution of
$\prm{b}=\prm{1} \Rightarrow
\prm{b'}=\prm{1}$.
\end{enumerate}
It is straightforward to observe that $\phi$ is a solution 
of the equations in (ii), by using Lemma \ref{lem:bangconditionlemma}.
Therefore, $\phi$ is a solution of $\Bang(\lift{M})$.

Now suppose the converse and let $u$ be 
a (parameterized) bang subterm 
with the critical parameter $\prm{b}$ in $\lift{M}$.
Suppose also that $\phi(u)$ is a bang subterm of $\phi(\lift{M})$.
This means that $\phi(\prm{b})= 1$. Since $\phi$ is supposed to 
be a solution of $\Bang(\lift{M})$, $u$ must contain at most
one free variable, say $x^{\pa^{\prm{b'},\prm{c'}} F'}$.
Moreover, we have $\phi(\prm{b'})=1$, which means that
$\phi(x)$ has a bang type in $\phi(\lift{M})$. Therefore,
$\phi(\lift{M})$ satisfies the Bang condition (i).
 As $\phi$ satisfies the conditions in (ii) and $\phi(\prm{b})= 1$
we get that $\phi(u)$ satisfies the condition
in Lemma \ref{lem:bangconditionlemma}, hence by this Lemma we obtain
that $\phi(u)$ satisfies the Bang condition (ii).
\end{proof}

\section{Solving the constraints}

Having described a way to collect a set of constraints from 
a given system F term,
there just remains to give a fast algorithm to solve them.
 Our method proceeds as follows: first solve the
boolean constraints, which corresponds to determine which $\bs$-boxes
are necessary (in \ref{ss-boolean}), and then solve the integer constraints, which corresponds 
to complete the decoration by finding a suitable
box structure (in \ref{ss-integer}).

\subsection{Solving boolean constraints}\label{ss-boolean}

We split $\Const(\lift{M})$ into three disjoint sets
$\Const^b(\lift{M})$, $\Const^i(\lift{M})$, $\Const^m(\lift{M})$:
\begin{enumerate}[$\bullet$]
\item A {\em boolean constraint} $\prm{s}\in\Const^b(\lift{M})$ consists
of only boolean parameters. $\prm{s}$ is of one of the following forms:\\

\begin{tabular}{llll}
$\prm{b_1} = \prm{b_2}$ & (in $\Ltype(\lift{M})$), \qquad \qquad & $\prm{b} = \prm{1}$ & (in $\Ltype(\lift{M})$),\\
$\prm{b} = \prm{0}$ & (in $\Bang(\lift{M})$), \qquad & $\prm{b} = \prm{1}\Rightarrow
\prm{b'} = \prm{1}$ &  (in $\Bang(\lift{M})$).
\end{tabular}\\

\item A {\em linear constraint} $\prm{s}\in\Const^i(\lift{M})$ deals
  with integer parameters only. A linear constraint
  $\prm{s}$ is  of one of the following forms:\\

\begin{tabular}{ll}
  $\prm{c_1} = \prm{c_2}$ & (in $\Ltype(\lift{M})$), \\
  $\prm{c} \geq \prm{0}$ & (in $\Ltype(\lift{M})$, $\Bra(\lift{M})$,
  $\Scope(\lift{M})$),\\
  $\prm{c} = \prm{0}$ & (in $\Ltype(\lift{M})$ and  $\Bra(\lift{M})$).
\end{tabular}\\

\item A {\em mixed constraint} $\prm{s}\in\Const^m(\lift{M})$ contains
  a boolean parameter and a linear combination and is of the following
  form:\\

\begin{tabular}{ll}
  $\prm{b} = \prm{1}\Rightarrow \prm{c} = \prm{0}$
  & (in $\Bang(\lift{M})$),\\
  $\prm{b} = \prm{1}\Rightarrow \prm{c} \geq \prm{1}$
  & (in $\Ltype(\lift{M})$ and $\Bang(\lift{M})$).
\end{tabular}
\end{enumerate}

We first try to find a solution of $\Const^b(\lift{M})$, and
then proceed to the other constraints. This does not cause loss of
generality,
because $\Const^b(\lift{M})$ admits a {\em minimal} solution 
whenever solvable.
Let us consider the set of instantiations on boolean parameters and the
extensional order $\leq$ on these maps: $\psi^b \leq \phi^b$ if for
any $\prm{b}$, $\psi^b(\prm{b}) \leq \phi^b(\prm{b})$.

\begin{lem}\label{minimalsolution}
There is a polynomial time algorithm to decide whether 
  $\Const^b(\lift{M})$ has a solution or not. Moreover,
the algorithm returns a minimal solution whenever there exists any.
\end{lem}
\begin{proof}
Our algorithm is based on the standard resolution procedure.
 Let $\bc :=\Const^b(\lift{M})$.  Apply repeatedly the following
  steps until reaching a fixpoint:
\begin{enumerate}[$\bullet$]
\item if $\prm{b_1}=\prm{b_2} \in \bc $ and $\prm{b_1}=\prm{i} \in \bc $
with $\prm{i}\in\{\prm{0},\prm{1}\}$, 
then let $\bc := \bc  \cup \{\prm{b_2}=\prm{i}\}$;
\item if $\prm{b_1}=\prm{b_2} \in \bc $ and $\prm{b_2}=\prm{i} \in \bc $
with $\prm{i}\in\{\prm{0},\prm{1}\}$, 
then let $\bc := \bc  \cup \{\prm{b_1}=\prm{i}\}$;
\item if $ (\prm{b}=\prm{1} \Rightarrow \prm{b'}=\prm{1})
\in \bc $ and $\prm{b}=\prm{1}
\in \bc $, then let $\bc := \bc  \cup \{\prm{b'}=\prm{1}\}$.
\end{enumerate}
 It is obvious that this can be done in a linear number of steps and
that the resulting system $\bc $ is equivalent to $\Const^b(\lift{M})$.

Now, if $\bc $ contains a pair of equations: 
$\prm{b}=\prm{0}, \prm{b}=\prm{1}$, then
it is inconsistent, and hence $\Const^b(\lift{M})$ does not 
have a solution. 
Otherwise, define the boolean instantiation
$\psi^{b}$ by
$$
\begin{array}{rcll}
\psi^{b}(\prm{b}) & := & 1 & \mbox{ if $\prm{b}=\prm{1} \in \bc $;}\\
 & := & 0 & \mbox{ otherwise.}
\end{array}
$$
It is clear that $\psi^{b}$ is a solution of $\bc $. In particular,
observe that any constraint of the form $(\prm{b}=\prm{1} \Rightarrow
\prm{b'}=\prm{1})$ in $\bc $ is satisfied by $\psi^{b}$. Moreover any
solution $\phi^{b}$ of $\bc $ satisfies $\psi^b \leq \phi^b$.
Therefore, $\psi^{b}$ is a minimal solution of
$\Const^b(\lift{M})$.
\end{proof}

\subsection{Solving integer constraints}\label{ss-integer}

When $\phi^{b}$ is a boolean instantiation, 
$\phi^{b}\Const^m(\lift{M})$ denotes
the set of linear constraints defined as follows:
\begin{enumerate}[$\bullet$]
\item  for any constraint of the form $(\prm{b} = \prm{1}\Rightarrow
\mathcal{I})$ in $\Const^m(\lift{M})$, where $\mathcal{I}$ is a linear (in)equation
(of the form $\prm{c} \geq \prm{1}$ or $\prm{c}= \prm{0}$),
$\mathcal{I}$ belongs to $\phi^{b}\Const^m(\lift{M})$
 if and only if $\phi^b (\prm{b})=1$.
\end{enumerate}
Then we clearly have:
\begin{enumerate}[(*)]
\item $(\phi^b, \phi^i)$ is a solution of
$\Const(\lift{M})$ if and only if $\phi^b$ is a solution of
$\Const^b(\lift{M})$ and $\phi^i$ is a solution of
$\phi^{b}\Const^m(\lift{M})\cup\Const^i(\lift{M})$.
\end{enumerate}
\begin{lem}\label{solutionwithminimalbooleans}
  $\Const(\lift{M})$ admits a solution if and only if it has a
  solution $\psi=(\psi^{b}, \psi^{i})$ such that $\psi^{b}$ is the
  minimal solution of $\Const^b(\lift{M})$.
\end{lem}
\begin{proof}
Suppose that   $\Const(\lift{M})$ admits a solution $(\phi^b, \phi^i)$.
Then by the previous Lemma, there is a minimal solution $\psi^b$
of $\Const^b(\lift{M})$. Since $\psi^b\leq \phi^b$, we have
  $\psi^{b}\Const^m(\lift{M})\subseteq 
\phi^{b}\Const^m(\lift{M})$. Since $\phi^i$
is a solution of $\phi^{b}\Const^m(\lift{M})\cup\Const^i(\lift{M})$
by (*) above, it is also a solution of 
$\psi^{b}\Const^m(\lift{M})\cup\Const^i(\lift{M})$.
This means that $(\psi^b, \phi^i)$ is a solution of 
$\Const(\lift{M})$.
\end{proof}


Coming back to the proof-net intuition, Lemma
\ref{solutionwithminimalbooleans} means that given a syntactic tree of
term there is a most general (minimal) way to place $\bs$-boxes (and
accordingly $\bs$ subtypes in types), that is to say: if there is a
\DLAL\ decoration for this tree then there is one with precisely this
minimal distribution of $\bs$-boxes.

Now notice that 
$\psi^{b}\Const^m(\lift{M})\cup\Const^i(\lift{M})$ is a linear
inequation system, for which a polynomial time procedure for searching
a rational solution is known (\cite{Kachian79,Karmarkar84}).

\begin{lem}\label{integersolution}
  $\psi^{b}\Const^m(\lift{M})\cup\Const^i(\lift{M})$ has a solution in
  $\QQ$ if and only if it has a solution in $\ZZ$.
\end{lem}

\begin{proof}
  Clearly the set of solutions is closed under multiplication by a
  positive integer.
\end{proof}

\begin{thm}\label{t-const}
  Let $M$ be a system F term. Then one can decide in time polynomial
  in the cardinality of $\Const(\lift{M})$ whether
  $\Const(\lift{M})$ admits a solution.
\end{thm}

\begin{proof}
  First decide if there is a solution of $\Const^b(\lift{M})$,
and if it exists, let $\psi^b$ be the minimal one 
(Lemma \ref{minimalsolution}).
Then apply the polynomial
time procedure to decide if
$\psi^b\Const^m(\lift{M})\cup\Const^i(\lift{M})$ admits a solution
in $\QQ$. If it does, then we also have an integer solution 
(Lemma \ref{integersolution}).
Otherwise, $\Const(\lift{M})$ is not solvable.
\end{proof}


By combining Theorems \ref{t-correct}, \ref{instantiationtheorem},
\ref{t-param} and \ref{t-const}, we conclude that 
the \DLAL\ typing problem (Problem \ref{problemtyping}) can be
solved in polynomial time:

\begin{thm}\label{t-main}
Given a system F term $M^T$, it is decidable in
time polynomial in the size of $M$ whether
there is a decoration $A$ of $T$ such that
$\vdash_{DLAL} M: A$.
\end{thm}

\section{Data-types and typing with domain specification}\label{sect:domainspecification}

\subsection{Data-types}
 Now that we have a type inference procedure, we can reexamine
the data-types in system F and the corresponding types in \DLAL.

Consider for instance the system F type for unary integers:
$$N_F=\forall \al. (\al \rightarrow \al)\rightarrow (\al \rightarrow
\al) \; .$$
We denote by $\underline{k}$ the
Church integer for $k$.

If we apply the type inference procedure to the Church integer $\underline{2}$,
we obtain the following family of parameterized types with constraints
as result:
$$
\left\lbrace\begin{array}{l}
A=\pa^{\prm{n_1}}  \forall \al.\pa^{\prm{n_2}}[
 \pa^{\prm{b_3},\prm{n_3}}(\pa^{\prm{b_4},\prm{n_4}}\al \fm \pa^{\prm{n_5}}\al)
\fm \pa^{\prm{n_6}}(\pa^{\prm{b_7},\prm{n_7}}\al \fm \pa^{\prm{n_8}}\al)
]\\
 \prm{b_3}=\prm{1},\; \prm{b_4}=\prm{b_7}=\prm{0}  \\ 
 \prm{n_4}=\prm{n_5}, \;  \prm{n_7}=\prm{n_8}, \;\\
 \prm{n_3}+  \prm{n_4}=\prm{n_6} +\prm{n_7} \\ 
\prm{n_7}\geq \prm{n_4}\\
 \prm{n_i}\geq \prm{0}, \; \prm{n_j}\geq \prm{b_j} \mbox{ for $1 \leq i \leq 8$ and $j=3.$}\\
\end{array}\right.
$$

It is easy to check that conversely, any solution to this system gives
a type suitable for all Church integers.  We denote by
$\mathcal{N}(A)$ this set of constraints. If $D$ is a free bang
decoration of $N_F$, we define $\mathcal{N}(D)=
\mathcal{N}(D^{\circ})$.

Observe that the type $N_{DLAL}= \forall \al. (\al\fm \al)\fli \pa
(\al\fm \al)$ is obtained by a solution of this system
 ($\phi(\prm{n_3})=\phi(\prm{n_6})=1$, $\phi(\prm{n_i})=0$ for $i\neq 3,6$,
 $\phi(\prm{b_3})=1$, $\phi(\prm{b_4})=\phi(\prm{b_7})=0$) 
 but it is not
the only one. For instance the following types are also suitable \DLAL\ types for
Church integers:
\begin{enumerate}[$\bullet$]
\item $N'_{DLAL}=\forall \al. (\al\fm \al)\fli (\pa \al\fm
\pa \al)$,

obtained with $\phi_0$ defined as the previous $\phi$, but for
$\phi_0(\prm{n_6})=0$, $\phi_0(\prm{n_7})=\phi_0(\prm{n_8})=1$.

\item  $\pa \forall \al. \pa^{2} (\pa \al\fm \pa \al)\fli
\pa (\pa^{3} \al\fm \pa^{3}\al)$,

obtained with $\phi_1(\prm{n_i})=1$ for $i=1, 4, 5, 6$;  $\phi_1(\prm{n_i})=3$ for $i=3, 7, 8$;
$\phi_1(\prm{n_2})=0$,  $\phi_1(\prm{b_3})=1$, $\phi_1(\prm{b_4})=\phi_1(\prm{b_7})=0$.
\end{enumerate}

 In the same way we can characterise the \DLAL\ types for the Church
representations of binary words, with a linear free decoration $A$
of the system F type $W_F$ and 
the following set of constraints
$\mathcal{W}(A)$.

{\small
$$
\left\lbrace\begin{array}{l}
    A=\pa^{\prm{n_1}}  \forall \al.\pa^{\prm{n_2}}[
    \pa^{\prm{b_3},\prm{n_3}}(\pa^{\prm{b_4},\prm{n_4}}\al \fm \pa^{\prm{n_5}}\al)
    \fm \pa^{\prm{n_6}}[
    \pa^{\prm{b_7},\prm{n_7}}(\pa^{\prm{b_8},\prm{n_8}}\al \fm \pa^{\prm{n_9}}\al)
    \fm
    \pa^{\prm{n_{10}}}(\pa^{\prm{b_{11}},\prm{n_{11}}}\al \fm \pa^{\prm{n_{12}}}\al) ]]\\
    \prm{b_3}=\prm{b_7}=\prm{1}, \quad \prm{b_4}=\prm{b_8}=\prm{b_{11}}=\prm{0}\; \\
    \prm{n_4}=\prm{n_5}, \; \prm{n_8}=\prm{n_9},\; \prm{n_{11}}=\prm{n_{12}},\\ 
    \prm{n_3}+\prm{n_4} =  \prm{n_6}+\prm{n_7} + \prm{n_8}, \;\\
    \prm{n_7}+\prm{n_8} =  \prm{n_{10}}+\prm{n_{11}}, \;\\
    \prm{n_{11}}\geq \prm{n_8}\\
    \prm{n_{11}}\geq \prm{n_4}\\
    \prm{n_i}\geq \prm{0}, \; \prm{n_j}\geq \prm{b_j} \mbox{ for $1 \leq i \leq 12$ and  $j=3,7.$}\\ 
\end{array}\right.
$$
}
\subsection{Typing with domain specification}\label{subsection:domainspecification}
Actually the \DLAL\ typability of a term \linebreak $M^{W_F\rightarrow
  W_F}$ of system F is \textit{not} sufficient to ensure that $M$ is
Ptime computable. To illustrate this point, we consider for simplicity
unary Church integers
and terms of type $N_F \rightarrow N_F$. 
 Observe that the following term of system F has type
$N_F \rightarrow N_F$ and represents the exponentiation function ($2^n$)
over unary integers:
$$ exp=\la n. \La \beta. (n \; \beta \rightarrow \beta) (\underline{2}\; \beta).$$
 Thus the term $exp$ does not represent a Ptime function but\ldots  it is typable in \DLAL,
with for instance the type:
$$ \forall \al. [(\al \fli \pa \al)\fm (\al \fli \pa \al)] \fm 
\forall \beta. [(\beta \fm \beta)\fli \pa (\beta \fm \beta)].$$ The trick
here is that this \DLAL\ type does not allow the term $exp$ to be applied
to all Church integers. Indeed the only closed terms of type $\forall
\al. [(\al \fli \pa \al)\fm (\al \fli \pa \al)]$ are $\underline{0}$
and $\underline{1}$. So we do obtain a Ptime term but over a
restricted, finite domain\ldots

 In general we are therefore not just interested in mere typability
but in typability with meaningful types. Indeed we generally want the
terms to be typable in \DLAL\ in such a way that they can be applied to
arguments of certain data-types (unary integers, lists\dots). This can
be enforced by adding some specification about the domain of the
function.

 Let $M$ be a system F term of type $T$. We call a \textit{domain specification} of $M$
a list $Dom=\langle (x_1,s_1), \dots, (x_k,s_k)\rangle$ such that for each $i$:
\begin{enumerate}[$\bullet$]
\item $x_i$ is a bound variable of $M$,
\item $s_i \in \{\mathbf{N}, \mathbf{W}\}$,
\item if $s_i= \mathbf{N}$ (resp.  $s_i= \mathbf{W}$), then $x_i$ is
of type $N_F$ (resp. $W_F$) in $M$.
\end{enumerate}

 For instance for the previous example of term $exp$ we can take 
$Dom=\{(n, \mathbf{N})\}$.

 Here we restrict to  $N_F$, $W_F$ for simplicity, but this definition could
be extended to other data-types of system F such as lists, binary trees\ldots

  Now we consider the free decoration $\lift{M}$. Let
$\mathsf{DomConst}(\lift{M}, Dom)$ be the union of $\mathcal{N}(D_i)$
(resp.  $\mathcal{W}(D_i)$) for all bound variables $x_i^{D_i}$ such
that $(x_i, \mathbf{N})$ (resp.  $(x_i, \mathbf{W})$) is in $Dom$.
 
 Finding a \DLAL\ type for $M$ such that, in the resulting \DLAL\ typed
term, each $x_i$ from $Dom$ can be instantiated with a Church integer
or binary word, is thus equivalent to finding a solution of
$\Const(\lift{M})$ which also satisfies $\mathsf{DomConst}(\lift{M},
Dom) $. In the previous example of $exp$ and the domain specification
$Dom$, there is not any such solution.

We have:

\begin{thm}
  Let $M$ be a System F term and $Dom$ be a domain specification. One can decide in time polynomial
  in the cardinality of $\Const(\lift{M})\cup \mathsf{DomConst}(\lift{M}, Dom) $ whether
  it  admits a solution.
\end{thm}
\begin{proof}
It is sufficient to observe that the constraints in $\mathcal{N}(D)$ (where $D$
is a decoration of $N_F$ or $W_F$)
also satisfy the properties used to prove Lemma \ref{minimalsolution},
Lemma \ref{solutionwithminimalbooleans} and Lemma
\ref{integersolution}.

 Note that for Lemma \ref{integersolution} for
instance the argument would not be valid anymore (at least in an
obvious way) if we added constraints of the form $\prm{n}=1$ or
$\prm{n}\leq 1$.
\end{proof}

%
 
 Therefore one can perform \DLAL\ decoration for system F terms in
polynomial time even with domain specification.

\section{Implementation and examples}\label{l-implement}

\subsection{Overview}

 We designed an implementation of the type inference
algorithm. The program is written in functional CAML and is quite
concise (less than 1500 lines). A running program not only shows the
actual feasibility of our method, but is also  a great facility for
building examples, and thus might allow for a finer study of the
algorithm.

Data types as well as functions closely follow the previous
description of the algorithm: writing the program in such a way tends
to minimise the number of bugs, and speaks up for the robustness of
the whole proof development.

The program consists of several successive parts:
\begin{enumerate}[(1)]
\item{Parsing phase: turns the input text into a concrete syntax
    tree. The input is a system F typing judgement, in a Church style
    syntax with type annotations at the binders. It is changed into
    the de Bruijn notation, and parameterized with fresh
    parameters. Finally, the abstract tree is decorated with
    parameterized types at each node.}
\item{Constraints generation: performs explorations on the tree
    and generates the boolean, linear and mixed constraints.}
\item{Boolean constraints resolution: gives the minimal solution of the
boolean constraints, or answers negatively if
    the set admits no solution.}
\item{Constraints printing: builds the final set of linear constraints.}
\end{enumerate}

We use a solver employing the simplex algorithm to solve the linear constraints. It runs in time
$O(2^n)$, which comes in contrast with the previous result of
polynomial time solving, but has proven to be the best in practice.

We now have to define the objective function that we will give to the
solver. Basically, to minimize the resulting complexity bound, we should
have an objective function which minimizes the nesting depth of the
boxes of the typed term. To achieve this, we would have to minimize the
maximum of the sums of door parameters from the root to each node (this corresponds to
the depth of the proof-net, which yields the bound of Theorem \ref{DLALfundamentalproperties}).
This clearly calls for a minimax objective function.
Unfortunately, this does not fit into the linear programming setting
that we are currently using: our objective function can only be a
ponderated sum of variables.

 So, we chose to simply put as objective
function the sum of door parameters. A little trick is needed in order
to handle the case of variables which are not of positive domain, and
could lead to the absence of an optimal solution. Once this special
case is handled, the solver always gives sensible results in practice.

The program, together with some examples, is  available at:

 \url{http://www-lipn.univ-paris13.fr/~atassi/ }



\subsection{Two examples: reversing of list and predecessor}

\subsubsection{List reversing.}

Let us consider the reversing function on binary words. It can be
defined by a single higher-order iteration on the type $W_F$, with the
untyped term \linebreak $\la w. \la so . \la si. (w) \; step_0 \; step_1 \;
base$, with :
\begin{enumerate}[$\bullet$]
\item base term: $base= \la z.z$,
\item step terms: $step_0= \la a.\la x.(a) (so) x$, $step_1= \la a.\lambda x.(a) (si) x.$
\end{enumerate}
 We obtain as system F term the following one, denoted \textbf{rev}:
$$
\begin{array}{l}
  \lambda l^{W}.\Lambda \beta.\lambda so^{\beta \rightarrow \beta}.\lambda si^{\beta \rightarrow \beta}. (l \; (\beta \rightarrow \beta))\\ 
   \; \; \; \; \lambda a^{\beta \rightarrow \beta}.\lambda x^{\beta}.(a) (so) x\\
   \; \; \; \; \lambda a^{\beta \rightarrow \beta}.
   \lambda x^{\beta}.(a) (si) x \; \lambda z^{\beta}.z \; . 
 \end{array}
 $$
 As discussed
in Section \ref{subsection:domainspecification} to obtain a meaningful typing 
we need to force
the domain of the term to be that of binary words. For that a simple
way is to apply the term to a particular argument, for instance:
$ \Lambda \alpha.\lambda so^{\alpha \rightarrow \alpha}.\lambda si^{\alpha \rightarrow \alpha}.\lambda x^{\alpha}.(si) (so) (si) (so) x, $
representing the word \textbf{\underline{1010}}.
Since \textbf{rev} involves higher-order functionals and polymorphism, 
it is not so straightforward to tell,
just by looking at the term structure, 
whether it works in polynomial time or not.

Given \textbf{rev(\underline{1010})} as input (coded by ASCII characters),
our program produces 
200 (in)equations on 76 variables. After constraint solving,
we obtain the result:
$$
\begin{array}{l}
 (\lambda l^{W}.
   \Lambda \beta . \lambda so^{\bs (\beta \llto \beta)} . \lambda si^{\bs (\beta \llto \beta)}.\\
   \; \; \; \; \; \; \; \; \pa (\pad ((l \; (\beta \llto \beta)) \\
   \; \; \; \; \; \; \; \; \pa \lambda a^{\beta \llto \beta}.\lambda x^{\beta}.(a) (\pad so) x \\
   \; \; \; \; \; \; \; \; \pa \lambda a^{\beta \llto \beta}.\lambda x^{\beta}.(a) (\pad si) x)\\
   \; \; \; \; \; \; \; \; \lambda z^{\beta}.z) \;\\
   \Lambda \alpha.\lambda so^{\bs\alpha \rightarrow \alpha}.\lambda si^{\alpha \rightarrow \alpha}.\pa \lambda x^{\alpha}.(\pad si) (\pad so) (\pad si) (\pad so)\;. x \;.
 \end{array}
$$
 It corresponds to the natural depth-1 typing of the term \textbf{rev}, with
conclusion type $W_{DLAL}\fm W_{DLAL}$. 
 The solution ensures polynomial time termination, and 
in fact its depth guarantees normalization in a quadratic
number of $\beta$-reduction steps.

\subsubsection{Predecessor on unary integers.}

 We now turn to another example which illustrates the use of polymorphism:
the predecessor function on unary integers. 

 We consider a slight simplification of the term given by Asperti
(\cite{Asperti98}). The
 simplification is not needed for typability, but is just chosen to
facilitate readability. 

 For that we consider:
\begin{enumerate}[$\bullet$]
\item  pairs represented in the following way:
$<P, Q> \; : \; \lambda z.(z) \; P
\; Q$,
\item terms for projection and an application combinator for pairs:
 $$\begin{array}{rcl}
 fst &=& \lambda x.\lambda y.x, \\
 snd &=& \lambda x.\lambda y.y,\\
 appl &=& \lambda x.\lambda y.(x) y\; .
\end{array}$$
\end{enumerate}

We will do an iteration on type $N_F$, with:
\begin{enumerate}[$\bullet$]
\item base term:  $<I,x>$ (where $I = \lambda
x.x$),
\item step term:   $\lambda p.<f,(p) \; appl>$.
\end{enumerate}

 The untyped term will then be $\lambda n.((n) \; step \; base) \; snd$.

Let us specify the system F typing of the subterms:
\begin{enumerate}[$\bullet$]
\item $<P,Q> = \lambda z^{(\beta \rightarrow \beta) \rightarrow (\beta
\rightarrow \beta)}.(z) \; P^{\beta \rightarrow \beta} \; Q^{\beta} :
((\beta \rightarrow \beta) \rightarrow (\beta \rightarrow \beta))
\rightarrow \beta$,
\item $snd, appl : (\beta \rightarrow \beta) \rightarrow \beta \rightarrow
\beta$,
\item $step = \lambda p^{((\beta \rightarrow \beta) \rightarrow (\beta
\rightarrow \beta)) \rightarrow \beta}.\lambda z^{(\beta \rightarrow
\beta) \rightarrow (\beta \rightarrow \beta)}.(z) \; f^{\beta
\rightarrow \beta} \; (p) \; appl^{(\beta \rightarrow \beta)
\rightarrow (\beta \rightarrow \beta)} \; :$

 $(((\beta \rightarrow
\beta) \rightarrow (\beta \rightarrow \beta)) \rightarrow \beta)
\rightarrow (((\beta \rightarrow \beta) \rightarrow (\beta \rightarrow
\beta)) \rightarrow \beta) $,
\item $base = \lambda z^{(\beta \rightarrow \beta) \rightarrow (\beta
\rightarrow \beta)}.(z) \; I^{\beta \rightarrow \beta} \; x : ((\beta
\rightarrow \beta) \rightarrow (\beta \rightarrow \beta)) \rightarrow
\beta \;.$
\end{enumerate}

The overall F-typed term for predecessor, denoted \textbf{pred} is thus:
\begin{align*}
& \lambda n^{\forall \alpha.(\alpha \rightarrow \alpha) \rightarrow
(\alpha \rightarrow \alpha)}.\Lambda \beta.\lambda f^{\beta
\rightarrow \beta}.\lambda x^{\beta}.\\
&(\\
&\qquad (n \; ((\beta \rightarrow \beta) \rightarrow (\beta \rightarrow \beta)) \rightarrow \beta)\\
& \qquad \lambda p^{((\beta \rightarrow \beta) \rightarrow (\beta \rightarrow
\beta)) \rightarrow \beta}.\lambda z^{(\beta \rightarrow \beta)
\rightarrow (\beta \rightarrow \beta)}.(z) \; f^{\beta \rightarrow
\beta} \; (p) \; appl^{(\beta \rightarrow \beta) \rightarrow (\beta
\rightarrow \beta)}\\
& \qquad \lambda z^{(\beta \rightarrow \beta)
\rightarrow (\beta \rightarrow \beta)}.(z) \; \lambda a^{\beta}.a \; x)\\
&)\\
& \lambda x^{\beta \rightarrow \beta}.\lambda y^{\beta}.y \;.
\end{align*}

Observe that this term is linear (as Asperti's original one).  Again,
to force a meaningful typing we apply the term \textbf{pred} to a
Church integer argument, here the integer
$\underline{2}$.

The program then produces 220 equations, for 130 parameters. The
solver produces a solution, yielding the following type for the
subterm \textbf{pred}:
$$ (\forall \al. (\al \fm \al) \fli \pa (\al \fm \al)) \fm (\forall
\al. (\al \fm \al) \fli \pa (\al \fm \al)), $$ which corresponds to the
$N_{DLAL} \fm N_{DLAL}$ type.

\subsection{Experiments with larger examples: polynomials}

 In order to test our type inference program with larger examples it
is interesting to consider a family of system F terms of increasing
size. The family of terms representing polynomial functions  over unary integers
is a natural candidate for this goal, since in particular it is
important for the encoding of polynomial time Turing machines
in the system (\cite{AspertiRoversi02,BaillotTerui04}).

 Therefore we wrote a CAML program which given a polynomial $P$
outputs a system F term representing $P$ and with type 
$N_F \rightarrow N_F$, that can then be fed to the \DLAL\ type inference program.

 There is however a subtlety that needs to be stressed. Recall that in
order to represent polynomial functions in \LAL\ or \DLAL\ with suitable
types it is necessary to use type c\oe{}rcions
(\cite{AspertiRoversi02,BaillotTerui04}). These c\oe{}rcions are needed
just for typing reasons, and not for computational ones. However, if
we consider the system F terms underlying the \LAL\ or \DLAL\ terms for
polynomials the c\oe{}rcions are still present and correspond to explicit
subterms.

 So if we want our system F terms representing polynomials to be
typable in \DLAL\ we need to anticipate on the need for
c\oe{}rcions. Therefore our program generating system F terms for
polynomials is guided by the encoding of polynomials in \DLAL, in
particular it takes into account the placement of subterms for
c\oe{}rcions (even if the terms are not yet typed with modalities during
this phase). It should be stressed that this increases considerably
the size of the resulting term: in practice inside the resulting term
the subpart accounting for the management of c\oe{}rcions is larger than
the subpart performing a computational task\dots This makes however a
good test for our type inference program, since the typing is not
trivial and will put into use a large number of parameters and
constraints.

 In the following we will:
\begin{enumerate}[$\bullet$]
\item describe the encoding of polynomials used,
\item report on experiments of our  type inference  program on
terms of this family. 
\end{enumerate}

\textbf{Encoding of polynomials.}

 We recall from \cite{BaillotTerui04} the rules for c\oe{}rcions on type
$N_{DLAL}$ derivable in \DLAL: 
\begin{center}
\begin{tabular}{c}
 {\infer[\mbox{(Coerc-1)}]{; m:N_{DLAL}, \pa \Delta \vdash \coo{t}:\pa A}{ n:N_{DLAL}; \Delta \vdash t :A}}\\[1em] 
 {\infer[\mbox{(Coerc-2)}]{\Gamma ; m: N_{DLAL}, \Delta \vdash \cot{t}:A}{\Gamma ; n: \pa N_{DLAL}, \Delta \vdash t:A}}
\end{tabular}
 \end{center}
where $\coo{.}$ and $\cot{.}$ are contexts, which contain as free variables some variables of the
environments:
\begin{eqnarray*}
\coo{x}&=& (m (\la g. \la p. (g \;(\succ\; p))))(\la n.x) \ci{0}, \\
\cot{x}&=& (\la n. x)(m \; \succ \; \ci{0})\;.
\end{eqnarray*}
Here $succ$ is the usual term for successor.

Similarly we define the term $coerc= \la n. (n) \; \succ \; \underline{0}$, which
can be given in \DLAL\ any type $N_{DLAL} \fm \pa^{k} N_{DLAL}$ with $k \geq 0$.
  
 Multiplication can be represented by the (untyped) term $mult=\la n. \la m. u$ with 
$u= ((m) \;
\lambda k.\lambda f.\lambda x.(n) \: f \: (k) \: f \: x) \:
\underline{0}$. It can be given in \DLAL\ the type  $N_{DLAL} \fli N_{DLAL} \fm \pa N_{DLAL}$.

Now, in order to give multiplication any type:  $\pa^{k}N_{DLAL} \fm \pa^k N_{DLAL} \fm \pa^{k + 2} N_{DLAL}$
with $k \geq 0$ we can use c\oe{}rcions:
\begin{prooftree}

\AxiomC{...}
\UnaryInfC{$ n : N_{DLAL} ; m : N_{DLAL} \vdash u : \pa N_{DLAL} $}
\RightLabel{Coerc-1}
\UnaryInfC{$ ; n_2 : N_{DLAL}, m : \pa N_{DLAL} \vdash C_1[u] : \pa^{2} N_{DLAL} $}
\RightLabel{Coerc-2}
\UnaryInfC{$ ; n_2 : N_{DLAL}, m_2 : N_{DLAL} \vdash C_2[C_1[u]] : \pa^{2} N_{DLAL} $}
\RightLabel{$\pa$-i $\times k$}
\UnaryInfC{$; n_2 : \pa^k N_{DLAL}, m_2 : \pa^k N_{DLAL} \vdash C_2[C_1[u]] : \pa^{k + 2} N_{DLAL} $}
\end{prooftree}

 Note that there is here a small abuse of notation as now the free
variable of $C_1[.]$ is called $n_2$ (similarly for $C_2[.]$).

We will associate to each polynomial $P$ of $\NN[X]$ a system F term $t_P$
of type $N_F \rightarrow N_F$ representing it, and which is typable in \DLAL.
We first describe the encoding of monomials.

We define  the term $t_{X^n}$ by induction on $n$:
 $$t_{X^0}=\la x. \underline{1} \; , \quad t_{X^1}=\la x. x \; , \quad t_{X^{n+1}}=  \lambda x.C_1[(\la n_2. \la m_2.C_2[C_1[u]]) \; (t_{X^n}) \; x \; (coerc) \; x],$$ for $n\geq 1$.

 The term $t_{X^n}$ can be given in \DLAL\ the type $N_{DLAL} \fm \pa^{4n} N_{DLAL}$. Actually
a better encoding of monomials could be given, with a lower depth, but we stick
here to this one for simplicity.
To show that  $t_{X^n}$ can be typed with $N_{DLAL} \fm \pa^{4n} N_{DLAL}$ note that
it is easy to observe for
$t_{X^0}$ and $t_{X^1}$, and supposing it for $t_{X^n}$ we get
for $t_{X^{n+1}}$ ($N$ in this derivation stands for $N_{DLAL}$): 

{\tiny
\begin{prooftree}
\AxiomC{\ldots}
\RightLabel{ $\pa\mbox{ i}  \times 4n $}
\UnaryInfC{$\vdash \la n_2. \la m_2.C_2[C_1[u]] : \pa^{4n} N\fm \pa^{4n} N\fm \pa^{4n + 2} N$}
\AxiomC{\ldots}
\UnaryInfC{$;x_1 : N \vdash (t_{X^n})  x_1 : \pa^{4n} N $}
\BinaryInfC{$;x_1 : N \vdash  (\la n_2. \la m_2.C_2[C_1[u]]) \; (t_{X^n}) \; x_1 : \pa^{4n} N \fm \pa^{4n + 2} N$}
\AxiomC{\ldots}
\UnaryInfC{$;x_2 : N \vdash (coerc) \; x_2 : \pa^{4n} N $}
\BinaryInfC{$;x_1 : N, x_2 : N \vdash (\la n_2. \la m_2.C_2[C_1[u]]) \; (t_{X^n}) \; x_1 \; (coerc) \; x_2 : \pa^{4n + 2} N$}
\RightLabel{$\pa$i}
\UnaryInfC{$x_1 :  N, x_2 :  N ;\vdash (\la n_2. \la m_2.C_2[C_1[u]]) \; (t_{X^n}) \; x_1 \; (coerc) \; x_2 : \pa^{4n + 3} N$}
\RightLabel{contr}
\UnaryInfC{$x :  N ; \vdash ((\la n_2. \la m_2.C_2[C_1[u]]) \; (t_{X^n}) \; x \; (coerc) \; x : \pa^{4n + 3} N$}
\RightLabel{coerc-1}
\UnaryInfC{$;x : N \vdash C_1[(\la n_2. \la m_2.C_2[C_1[u]]) \; (t_{X^n}) \; x \; (coerc) \; x] : \pa^{4n + 4} N$}
\RightLabel{$\fm$i}
\UnaryInfC{$\vdash \lambda x.C_1[(\la n_2. \la m_2.C_2[C_1[u]]) \; (t_{X^n}) \; x \; (coerc) \; x] : N \fm \pa^{4(n+1) } N$}
\end{prooftree}
}

 Now, once $t_{X^{n}}$ has been defined it is easy to represent
monomials with coefficient, $\alpha_n X^n$ and, using the term for
addition and c\oe{}rcions again, arbitrary polynomials: $\sum\limits _{i
= 1}^{j} \alpha_{n_i}X^{n_i}$.

  \textbf{Experiments of type
inference.}

We wrote a small program implementing this encoding, which, given a
polynomial, produces a system F term representing it. Then we used it
to test our \DLAL\ type inference program. We give the results of the
experiments on a few examples, in the array of Figure
\ref{arraytypeinference}, where $\mathbf{t_P}$ denotes the F term
representing a polynomial.
Again we stress that the large size of $\mathbf{t_P}$, even for small
polynomials, is due to the c\oe{}rcions (for instance the encoding of
$x^5$ without c\oe{}rcions produces a term of size 322 --- which is
not typable however) and to the fact that all types are written
explicitly in the term, since it is written in \textit{\`a la Church} style
syntax.

%
%
 In the array the following quantities are reported:
\begin{enumerate}[$\bullet$]
\item  the size of $\mathbf{t_P}$ is the number of symbols of the term;
\item  the column \textbf{\# Par} stands for the number
of parameters in the resulting parameterized term,
\item the time (in seconds) for generating the set of linear constraints is divided in two parts:
\begin{itemize}
\item \textbf{GEN} is the time taken by the program for parsing the
input, generating the whole constraints and solving the boolean part,
\item \textbf{SIMPL} is the time  taken to simplify the set of constraints (this is a preprocessing before using the solver).
\end{itemize}
\item \textbf{\# Cons} is the cardinality of the set of
of linear constraints generated by the program,
\item \textbf{Sol} is the time taken by the solver (LPsol) to solve the set of constraints.
\end{enumerate}

 Recall that $N'_{DLAL}=\forall \al. (\al \fm \al) \fli \pa \al \fm
\pa \al$. We think that the fact that we obtain a $N'_{DLAL}$ type
instead of $N_{DLAL}$ on the right-hand-side is not significative here: we
could force obtention of a $N_{DLAL}$ type instead by techniques similar to
that of domain specification of Section \ref{sect:domainspecification}
(adding a constraint of the form $p=0$).

Note that the type obtained is slightly smaller (containing fewer
$\pa$ and of smaller depth) than the one described above: we obtain
the type $N_{DLAL} \fm \pa^{4n-3} N'_{DLAL}$ for $t_{X^n}$, and it is
possible to check that this is indeed a suitable type in general.

\begin{figure*}

\begin{tabular}{|c|c|c|c|c|c|c|c|c|}
\hline
 $P=X^n$ & Size of $\mathbf{t_P}$ & \textbf{\# Par} &\textbf{GEN}& \textbf{SIMPL} & \textbf{\# CONS} & \textbf{Sol} & Type obtained & $4n$ \\
 \hline

$X^2$ &      380 &            520    &         0.1 &    0.1  &   844  &   0.0  &   $ N \fm \pa ^5 N'$  & 8\\
\hline
$X^3$    &  750  &           1009    &        0.3  &   0.7  &   1649 &   0.1 &  $ N \fm \pa ^{9} N'$  & 12\\ 
\hline
$X^4$   &   1120 &           1498    &        0.9  &   1.7  &   2454 &   0.2 & $ N \fm \pa ^{13} N'$  &16\\ 
\hline
$X^5$   &   1490 &           1987    &        1.9  &   3.3  &   3259 &   0.4 & $ N \fm \pa ^{17} N'$  &20\\ 
\hline
$X^6$  &     1860    &        2476   &         3.5 &    5.5 &    4064&    0.6 &$ N \fm \pa ^{21} N'$  &24\\ 
\hline
$X^7$  &    2230   &         2975    &        5.9  &   4.8  &   4869 &   0.8 &$ N \fm \pa ^{25} N'$  &28\\ 
\hline
$X^8$ &     2600  &          3454    &        9.0  &   6.9 &    5674 &   1.1 &$ N \fm \pa ^{29} N'$  &32\\ 
\hline
$X^9$  &     2970 &           3943    &        13.2 &   12.4 &   6479 &   1.5 &$ N \fm \pa ^{33} N'$  &36\\ 
\hline
$X^{10}$  &    3340 &            4432 &            18.5 &    21.0 &   7284 &    1.0 & $ N \fm \pa ^{37} N'$  & 40\\ 
\hline
$X^{16}$  &   5560 &            7336 &            86.6 &    80.5 &    12114  & 5.8 & $ N \fm \pa ^{61} N'$  &64\\ 
\hline
$X^{32}$  &   11480 &           15190 &           810.0 &  381.3 &  24994 &  30.5 &$ N \fm \pa ^{125} N'$  &128\\ 
\hline
\end{tabular}
 In the array $N$ (resp. $N'$) stands for $ N_{DLAL}$ (resp. $N'_{DLAL}$).

 \caption{Type inference for terms representing polynomials.}\label{arraytypeinference}
\end{figure*}

 Observe that on these examples the respective times needed for
generating the constraints and solving the boolean part (\textbf{GEN})
on the one hand, and for simplifying the linear constraints
(\textbf{SIMPL}) on the other, are comparable.  The time needed to
solve the linear constraints (\textbf{Sol}) is comparatively smaller.
 
 We also generated the system F terms representing the same polynomials but without
subterms for c\oe{}rcions, and noted with our program that type inference in \DLAL\ 
for these terms fails: c\oe{}rcions are indeed necessary.

 Even though the family of terms $t_{X^n}$ is a particular case, these
examples illustrate the fact that our algorithm is manageable with
lambda-terms of reasonable size, and gives results in a sensible time.

\section{Discussion and further work for the case of propositional \DLAL}

 It should be stressed that our method can be applied to type untyped
lambda-terms in propositional (quantifier-free) \DLAL. Indeed,
propositional \DLAL\ can naturally be seen as a subsystem of \DLAL. Given
an untyped term $t$, we can thus proceed in the following way (in the
lines of previous works for \EAL\ or \LAL\ like
\cite{CoppolaMartini01,Baillot02}):
\begin{enumerate}
\item search for the principal simple type  of $t$,
\item using the principal simple type derivation of $t$, search 
for a valid \DLAL\ decoration by using our method.
\end{enumerate}
 If we find a suitable decoration then it will give a derivation in
propositional \DLAL\ (simply because the underlying system F derivation
does not use quantification). It can be checked that this method is
complete (for instance by a simple adaptation of the argument in
\cite{Baillot04}): if the term is typable in propositional \DLAL, then
a suitable decoration of the principal simple type decoration will be
found.

 However, the bound on this procedure given by Theorem \ref{t-main} is
polynomial w.r.t. the \textit{size of the principal simple type derivation of
$t$}, and not w.r.t. to the \textit{size $|t|$ of the untyped term $t$ itself}.

Still, we strongly believe that our method can be adapted in order to
give an algorithm performing type inference in propositional \DLAL\ for
an untyped term $t$ in time polynomial in $|t|$.

 The starting point is that it is known that simple type inference can be
done in polynomial time by using a shared representation of types. If
one designs an algorithm performing together simple type inference and
decoration with parameters, one can presumably obtain, instead of a
\textit{free decoration} of $t$, a suitable decoration with a number of
parameters polynomial in $|t|$ (by taking advantage of the shared
representation of types) and a constraints system also polynomial in
$|t|$. Hence in the end type inference would be polynomial
w.r.t. $|t|$. We also believe that in this way we would obtain a
notion of principal propositional \DLAL\ type. This would be analogous
to the work of \cite{CoppolaRonchi03} for \EAL, but could give a single
principal type scheme instead of a finite family of principal type
schemes.

 However in the present paper we preferred to follow the approach
starting with a system F typed term in order to be able to consider
second-order \DLAL\ typing, which is more interesting for expressivity
reasons (propositional \DLAL\ is not complete for polynomial time
computation). The case of polynomial time type inference for
propositional \DLAL\ is left for future work.

\section{Conclusion}
 We showed that typing of system F terms in \DLAL\ can be performed in
a feasible way, by reducing typability to a constraints solving
problem and designing a resolution algorithm.  This demonstrates a
practical advantage of \DLAL\ over \LAL, while keeping the other
important properties. We illustrated the manageability of our algorithm
by implementing it in CAML and giving some examples of type inference.
Note that other typing features could still be
automatically inferred, like c\oe{}rcions (see \cite{Atassi05} for the
case of \EAL).

This work illustrates how Linear logic proof-net notions like boxes
can give rise to techniques effectively usable in type inference, even
with the strong boxing discipline of \DLAL, which extends previous
work on \EAL. We expect that some of these techniques could be adapted
to other variants of Linear logic, existing (like Soft linear logic)
or to be defined in the future.

\nocite{Terui01,Terui07}
 \bibliographystyle{alpha}

\end{document}